\documentclass[useAMS,usenatbib]{mn2e}

\title[Simulations of galaxies with an active potential]
{Simulations of spiral galaxies with an active potential: molecular cloud formation and gas dynamics}
\author[C. L. Dobbs, I. A. Bonnell]
{C. L. Dobbs$^1$\thanks{E-mail:
dobbs@astro.ex.ac.uk} \& I. A. Bonnell$^2$ \\
$^1$ School of Physics, University of Exeter, 
Stocker Road, Exeter, EX4 4QL \\
$^2$ SUPA, 
School of Physics and Astronomy, University of St Andrews, 
North Haugh, St Andrews, Fife, KY16 9SS}
\usepackage{amssymb}
\usepackage{amsmath}
\usepackage{graphicx}
\usepackage{epsfig}

\begin{document}
\date{\today}

\pagerange{\pageref{firstpage}--\pageref{lastpage}} \pubyear{0000}

\maketitle

\label{firstpage}

\begin{abstract}
We describe simulations of the response of a gaseous disc to an active spiral potential. 
The potential is derived from an N-body calculation and leads to a multi-armed time-evolving pattern. 
The gas forms long spiral arms typical of grand design galaxies,
although the spiral pattern is asymmetric. 
The primary difference from a  grand-design spiral galaxy, which has a consistent 2/4-armed pattern,
is that instead of passing through the spiral arms, gas generally falls into a developing potential minimum and is released only when the local minimum dissolves. In this case, the densest gas is coincident with the spiral potential, rather than offset as in the grand-design spirals.  We would therefore expect no offset between the spiral shock and star formation, and no obvious co-rotation radius.
Spurs which occur in grand-design spirals when large clumps are sheared off leaving the spiral arms, are rare in the active, time-evolving spiral reported here. Instead, large branches
are formed from spiral arms when the underlying spiral potential is dissolving due to the N-body dynamics. We find that the molecular cloud mass spectrum for the active potential is similar to that for clouds in grand design calculations, depending primarily on the ambient pressure rather than the nature of the potential. The largest molecular clouds occur when spiral arms collide, rather than by agglomeration within a spiral arm.
\end{abstract}


\section{Introduction}
A key objective to understanding star formation is to explain how molecular clouds and stars form in different types of galaxies. 
The presence of spiral shocks in grand design galaxies explains the predominance of young stars in spiral arms \citep{Roberts1969,Woodward1976,Cepa1990, Knapen1992,Knapen1996}, and can also account 
for the observed velocity dispersions in molecular clouds (\citealt{Zhang2001,Bonnell2006}; \citealt*{KKO2006}).
Otherwise, star formation is thought to occur through triggering by supernovae \citep{Mueller1976,Gerola1978}, spontaneously through gravitational, thermal or magnetic instabilities, or by turbulent compression (as reviewed in \citealt{Elmegreen1996}). 
 
From optical images, spiral galaxies can be classified as i) grand design, usually consisting of 2 symmetrical spiral arms, ii) multi-armed, with several asymmetric spiral arms or iii) flocculent, with multiple shorter arms.
Using the classification scheme of \citet{Elmegreen1987}, grand design galaxies are inclusive of multi-armed spiral galaxies, but we retain this terminology to distinguish between the two types.
Classical grand design structure is thought to originate from perturbations due to bars or companion galaxies which induce a density wave in the underlying stellar disk.
Flocculent (and multi-armed) structure develops from local gravitational instabilities, which are sheared into short, transient spiral arms.
Most flocculent galaxies are also found to have a weak density wave in the older stellar population visible in the K band \citep{Thornley1997,Grosbol1998}.

Several hydrodynamical simulations have investigated the response of the ISM to spiral density waves in grand design galaxies, by assuming a rigidly rotating spiral potential, usually with an isothermal medium \citep*{Patsis1997,Chak2003,Kim2002,Dobbs2006}. 
The spiral pattern in these simulations is assumed to be long-lasting, of at least several rotation periods. 
The formation of dense structures along the spiral arms are associated with molecular clouds \citep*{Kim2003,DBP2006,Dobbs2007} and the shearing of these features produces spurs perpendicular to the arms \citep*{Kim2002,Dobbs2006,Shetty2006}. 
For the purely hydrodynamic simulations \citep{Wada2004,DBP2006}, the development of molecular clouds and spurs is controlled by the dynamics of the shock.
Similar features are formed through gravitational instabilities \citep{Kim2002,Shetty2006}. 

The generation of turbulence has also been examined in simulations of galactic disks. \citet{Wada2002} show that gravitational instabilities combined with galactic rotation can drive turbulence, producing flocculent spiral structure. Supernovae are expected to be a major progenitor of turbulence \citep{deAvillez2005,Joung2006} in the ISM, whilst \citet{Piontek2005} suggest that MRI driven turbulence may be important in the outer parts of discs where supernovae are less frequent. An alternative possibility which has emerged recently is that spiral shocks may induce turbulence \citep{Zhang2001,Bonnell2006,KKO2006,Dobbs2007}. In particular, \citet{Bonnell2006} show the passage of gas through a spiral arm and the formation of molecular clouds which exhibit a $\sigma \propto r^{0.5}$ velocity dispersion law.

Realistic galactic potentials are not rigid, but time dependent. Even in grand design galaxies, the spiral pattern and pitch angle of the spiral arms may change within a few rotation periods \citep*{Merri2006}. 
For galaxies where the spiral structure arises through more localised gravitational instabilities, the number and shape of spiral arms are expected to be continually evolving. Numerous N-body simulations have investigated the spiral structure arising from such instabilities in a stellar disc. 
If the mass of the halo is sufficient, the bar instability is prevented and the  
resulting structure then depends on the Toomre $Q$ parameter.  
Generally gravitational instabilities produce flocculent and multi-armed spirals \citep*{Bottema2003,Li2005b}, unless an increase of $Q$ in the centre of the disk is induced to suppress $m>2$ modes \citep*{Thomasson1990}.
With $Q \sim 1-2$, multi-armed spirals with modes up to $m=6$ are formed \citep{Sellwood1984}.
Lowering the disk mass or raising the velocity dispersion increases $Q$ and produces a flocculent galaxy \citep{Elmegreen1993}. In all these cases, the gaseous component of the disc develops a much more pronounced spiral structure than present in the stellar component. Hydrodynamical simulations show that the gas becomes dense enough for star formation to occur in the spiral arms, providing the Toomre criterion \citep{Toomre1964} is satisfied ($Q \lesssim 1$ for gas \citep{Tasker2006} or gas and stars \citep{Li2005b}).
 
In this paper, we describe hydrodynamical simulations of multi-armed spiral galaxies, which use an active galactic potential. This potential is taken from an N-body simulation described in \citet{Sellwood1984}.
The potential has also recently been used for 2D gas simulations \citep{Clarke2006} with the grid code CMHOG, which highlight the dependence of the star formation rate on the changing spiral structure. Whilst the gas dynamics in the grand design simulations are driven by a density wave, the potential from the N-body simulation arises through transient gravitational instabilities in the disc stars. 
We describe the galactic structure, gas dynamics and the properties of molecular clouds which arise with the active potential, and compare with previous simulations where a rigid potential was adopted \citep{Dobbs2007}.
  
\section{Calculations}
We use the 3D smoothed particle hydrodynamics (SPH) code based on the version by
Benz \citep{Benz1990}. The smoothing length is allowed to vary with space and
time, with the constraint that the typical number of neighbours for each particle 
is kept near $N_{neigh} \thicksim50$.  
Artificial viscosity is included with the standard parameters $\alpha=1$
and $\beta=2$ \citep{Monaghan1985,Monaghan1992}. 

We describe the galactic potential and initial conditions for the simulations in Sections 2.1 and 2.2.
The 3 simulations performed are also summarised in Table~1. In all the simulations, 
self-gravity magnetic fields, heating, cooling or feedback from star formation are not included.
This paper focuses on the differences in the gas dynamics and structure resulting from the active potential. 

\begin{table*}
\begin{tabular}{c|c|c|c|c|c|c|c}
\hline
Run & No. of & Mass of cold & Mass of warm & Total mass & Average $\Sigma$ & Mass of H$_2$ after  \\
& particles & (100 K) gas (M$_{\odot}$) & ($10^4$ K) gas (M$_{\odot}$) & (M$_{\odot}$) & 
(M$_{\odot}$ pc$^{-2}$)  & 200 Myr (M$_{\odot}$) \\
 \hline
A & 27 million & 9 $\times 10^8$ &  $10^8$ & $10^9$ & 3.2 & 1.5 $\times 10^8$ \\
B & 27 million & 6 $\times 10^8$ & 6 $\times 10^8$ & 1.2 $\times 10^9$ &3.9 & 3.9 $\times 10^8$ \\
C & 4 million & 0 & 1.2 $\times 10^9$ & 1.2 $\times 10^9$ & 3.9 & - \\
\hline
\end{tabular}
\caption{Table showing the different simulations described, and the proportions of warm and cold gas for each.}
\end{table*}

\subsection{Galactic potential}
The gravitational potential for these calculations is taken from a previous N-body simulation \citep{Sellwood1984}. The N-body calculation includes both a disk and a rigid halo, the latter suppressing the formation of a bar in the disk. In the N-body simulation, the disk evolves to produce an asymmetric spiral pattern with several spiral arms. The distribution of particles in the calculation are shown at different times in Fig.~2 of \citet{Sellwood1984}.
The amplitude of the perturbation peaks after approximately 1 rotation period, measured at the midpoint of the disk, after which the intensity of the spiral pattern quickly decays. The spiral perturbation to the potential decays in the N-body simulation as the velocity dispersion increases, and the disc becomes stable (i.e. Q$>$1) against gravitational instabilities.   

The gravitational potential and accelerations were derived from the N-body simulation and used in previous results by \citet{Clarke2006}. Both quantities were calculated over a 2D cylindrical polar grid consisting of 90 x 128 points, after each time step in the N-body simulation (giving a total of 1500 grids). In this paper we just use the accelerations, which we interpolate both spatially and with time. Thus at a given time in the SPH calculation, we use the accelerations from the polar grids at the closest preceding and succeeding times.

Our simulations are 3D, so we also apply accelerations acting solely in the $z$ direction assuming an axisymmetric logarithmic disk potential, i.e.:
\begin{eqnarray}
F_z &=& \frac{-d\psi_{disk}}{dz} \quad \text{where}  \nonumber \\
\psi_{disk} &=& \frac{1}{2}v_0^2 \; log(r^2+R_c^2+(z/z_q)^2)
\end{eqnarray}
where $R_c=1$ kpc, $v_0=220$ km s$^{-1}$ and $z_q$=0.7 is a measure of the disc scale height.
The gravitational field is not fully consistent, since the $z$ component is not related to the spiral potential. Thus the generation of turbulence by 'flapping' motions of the spiral shocks \citep{KKO2006}, is not able to occur in these models, requiring either an enhanced $z$ component of the potential in the spiral arms, or self gravity. 
A fully consistent potential may also increase the density in the spiral arms, again due to the increased axial component of the potential there.

 \subsection{Initial distribution of gas}   
The particles are initially distributed uniformly across a disk with a scale height of 100 pc and maximum radius of 10 kpc.
We exclude particles from the inner 1 kpc radius of the disk, but as the simulation progresses, particles may enter this region.
The initial velocities of the particles assume the rotation curve given in \citet{Sellwood1984} (Equation~2), with a maximum rotational velocity of 220 km s$^{-1}$, and the particles are initially placed on circular orbits. We choose a length scale such that the maximum extent of the potential is 14 kpc (corresponding to r=7 in \citet{Sellwood1984}), although since the spiral perturbation does not extend to the whole of the disk in the N body simulation, we adopt a maximum radius of 10 kpc.
With this choice of length and velocity scales, the time for a simulation to complete the equivalent of the 1500 timesteps of the N-body calculation is 430 Myr. This corresponds to a time of t=4 on Fig. 2 of \citet{Sellwood1984}  (or equivalently 0.94 Gyr in \citealt{Clarke2006}). 

In addition to the rotational velocities, we also imposed a dispersion on the velocity distribution. The added dispersion is selected from a Gaussian distribution with a standard deviation of 2.5\% of the orbital speed. This was used for all the $x$, $y$, and $z$ velocities. The velocity dispersion in the $z$ direction maintains vertical equilibrium in the disk.
Turbulence is not explicitly included in the initial conditions of our simulations. However, as mentioned in the introduction, previous results have shown that spiral shocks induce a significant velocity dispersion in the gas. Thus the turbulence is generated self consistently in the simulation as gas is subject to spiral shocks, and the initial velocity dispersion is not very significant to the outcome of our results.

We perform 2 high resolution simulations, with 27 million particles (Table~1), where the resolution is approximately 40 M$_{\odot}$ per particle.
One calculation contains predominantly cold gas, with a much smaller fraction of warm gas, whilst the other uses equal phases of warm (10$^4$ K) and cold gas. 
The SPH particles are initially randomly assigned as warm or cold gas. We also perform a lower resolution calculation with just warm gas, and in all calculations the gas is isothermal. 

\subsection{Two phase medium}
In this present study, we neglect any heating or cooling between the 2 phases (so cold particles remain cold throughout the simulation). Although a full thermal treatment is preferable, a two phase medium is a first approximation to modeling cold and warm gas simultaneously, and allows a direct comparison with our previous calculations \citep{Dobbs2007}. 
Observations indicate both of these components are present in the ISM \citep{Heiles2003}, and thus it is important to include them in spiral shocks.
We assume that cold gas is pre-existing in the ISM (rather than the outcome of cooling from the warmer phase), as suggested by \citet*{Pringle2001}. Although if the cooling timescale of warm gas in the ISM is very short (i.e. a few Myr, e.g. \citealt{Audit2005,Heitsch2006,Vaz2007,Glover2007b}) cold clumps would quickly develop as warm gas enters the spiral arms, producing conditions in the spiral shock similar to these calculations. 

Although randomly positioned initially, the warm and cold phases become separated during the simulation. This separation is expected when gas passes through a spiral shock. In previous simulations \citep{Dobbs2007,Dobbs2008}, cold gas is compressed into well resolved clumps of order $10^6$ M$_{\odot}$. However phase separation is also apparent before the gas has time to pass through the shock. This is particularly noticeable at large radii in the calculations presented here, where the dynamical time is greatest. This is a numerical rather than physical artifact, and occurs due to a discontinuity in pressure between the warm and cold gas \citep{Price2007}. The separation of cold gas into clumps is accentuated by the shear velocities of the gas. When cold particles are adjacent to each other, a pressure imbalance between the two phases causes the cold gas particles to move closer together rather than move freely past each other. This pressure imbalance effectively suppresses mixing between the two phases of gas. 

Apart from in the inner regions of the disk, gas flowing into the spiral arms is situated in cold clumps surrounded by a warm medium. This distribution is nevertheless much more similar to the physical conditions in the ISM, where the cold gas is situated within higher density clumps, rather than randomly interspersed with the warm gas. However the distribution and spacing between these clumps is dependent on the resolution, with clumps typically containing 50-100 particles. Ideally we would wish to remove this numerical instability, ensuring clumps produced in the calculations are solely related to the spiral shocks. This should be feasible for future calculations, by way of an artificial thermal conductivity term \citep{Price2007}, which smoothes the pressure between adjacent gas particles of different temperatures.

\subsection{Calculation of molecular gas density and cooling of molecular gas}
We estimate the fraction of 
molecular gas using the equation for the rate of change of H$_2$ density from
\citet{Bergin2004},
\begin{equation}
\frac{dn(H_2)}{dt}=R_{gr}(T)n_p
n(H)-[\zeta_{cr}+\zeta_{diss}(N(H_2),A_V)]n(H_2).
\end{equation} 
The total number density is $n_p$, where $n_p=n(H)+2n(H_2)$, 
$N$ is the column density (of atomic or molecular hydrogen) and $T$ the 
temperature.
The photodissociation rate is
\begin{equation}
\zeta_{diss}=\zeta_{diss}(0)f(N(H_2),A_V),
\end{equation}
from \citet{Draine1996}. The constant $\zeta_{diss}(0)=4.17 \times 10^{-11}$ s$^{-1}$ is the photodissociation rate for unshielded H$_2$. The H$_2$ is assumed to be subject to a radiation field equivalent to the U.V. radiation emitted by B0 stars averaged over the plane of the Galactic disk. The function $f(N(H_2),A_V)$ takes into account self shielding of the H$_2$. This function includes the dust extinction ($A_V$), and the column density of the H$_2$, the latter calculated using the local density of H$_2$ and the scale height of the disk. The second destructive term is the cosmic ray ionization rate,
$\zeta_{cr}=6 \times 10^{-18}$ s$^{-1}$. 
The rate of formation on grains is $R_{gr}$, assuming an efficiency of
$S=0.3$. We do not explicitly change $S$ for the warm gas in
our calculations, but the density of the warm gas is too low to prevent immediate photodissociation of any H$_2$ formed \citep{DBP2006}. 

Although we do not include heating or cooling between the 2 phases, we do allow cooling
of the 100 K gas which becomes molecular. We use a simple polytropic equation of state
$P \propto \rho^{0.7}$ \citep{Larson2005} suggested for molecular clouds. We take 
only the molecular gas density for this equation of state, setting a maximum temperature of 100 K 
for gas which is fully atomic and a minimum temperature of 30 K for gas which is fully molecular.
It is not thought that the addition of molecular cooling has a significant effect on these simulations,
but may increase the density slightly in the shocks and prolong the time gas remains molecular. 
As the molecular gas dissociates, the gas heats back up (to a maximum temperature of 100 K) using the same equation of state.
\section{Results}
The primary driver for the evolution of the gas is the development  of spiral arms in the N-body potential. The main difference between the simulations presented here and those for a grand design galaxy is that the potential changes on timescales comparable to the gas crossing time of a spiral arm. Rather than gas passing through a spiral shock, the evolution and structure of the gas is determined by local changes in the potential minimum and collisions between spiral arms.
\subsection{Structure of the disc}
\begin{figure*}
\centerline{
\includegraphics[scale=0.33]{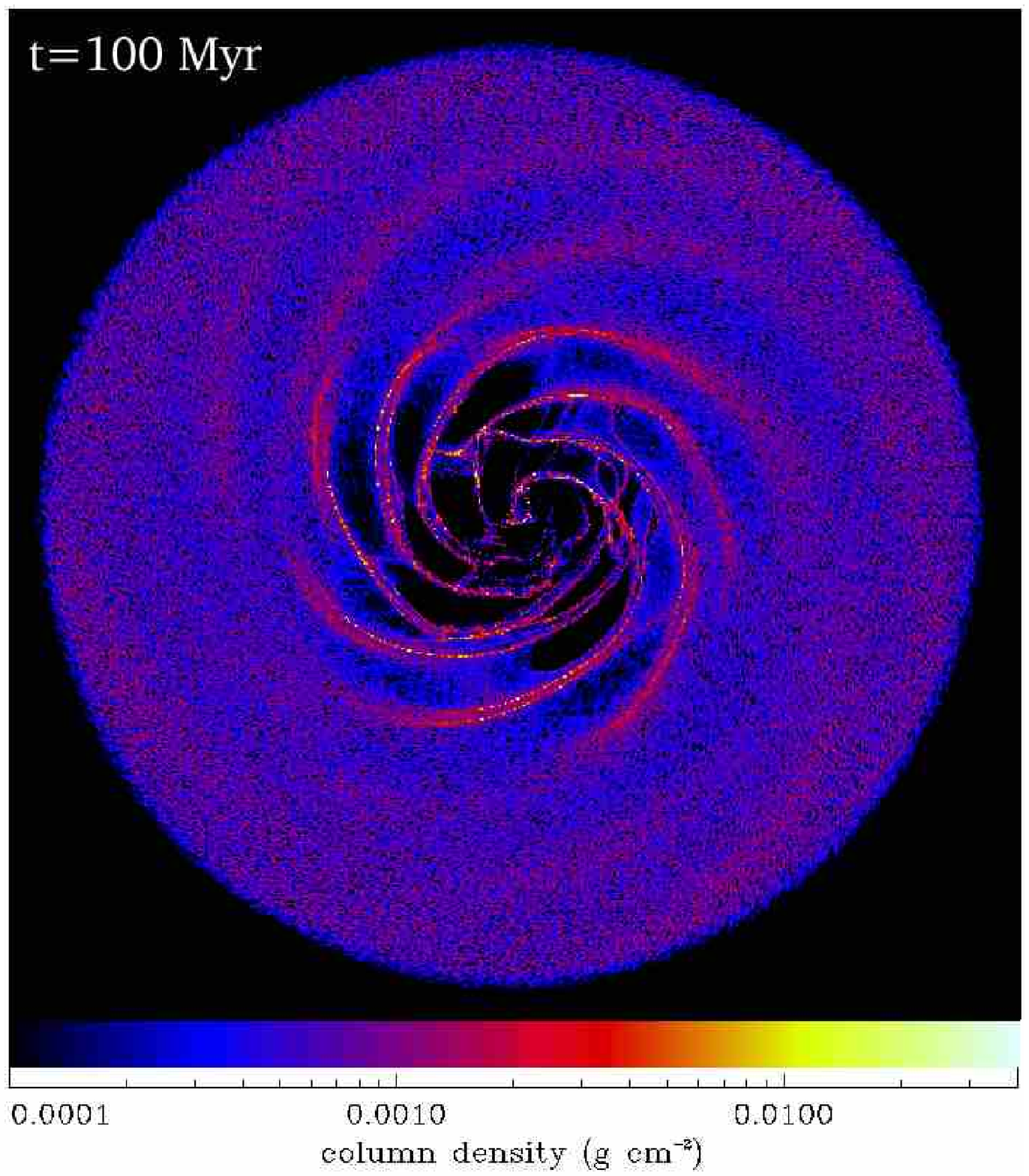}
\includegraphics[scale=0.33]{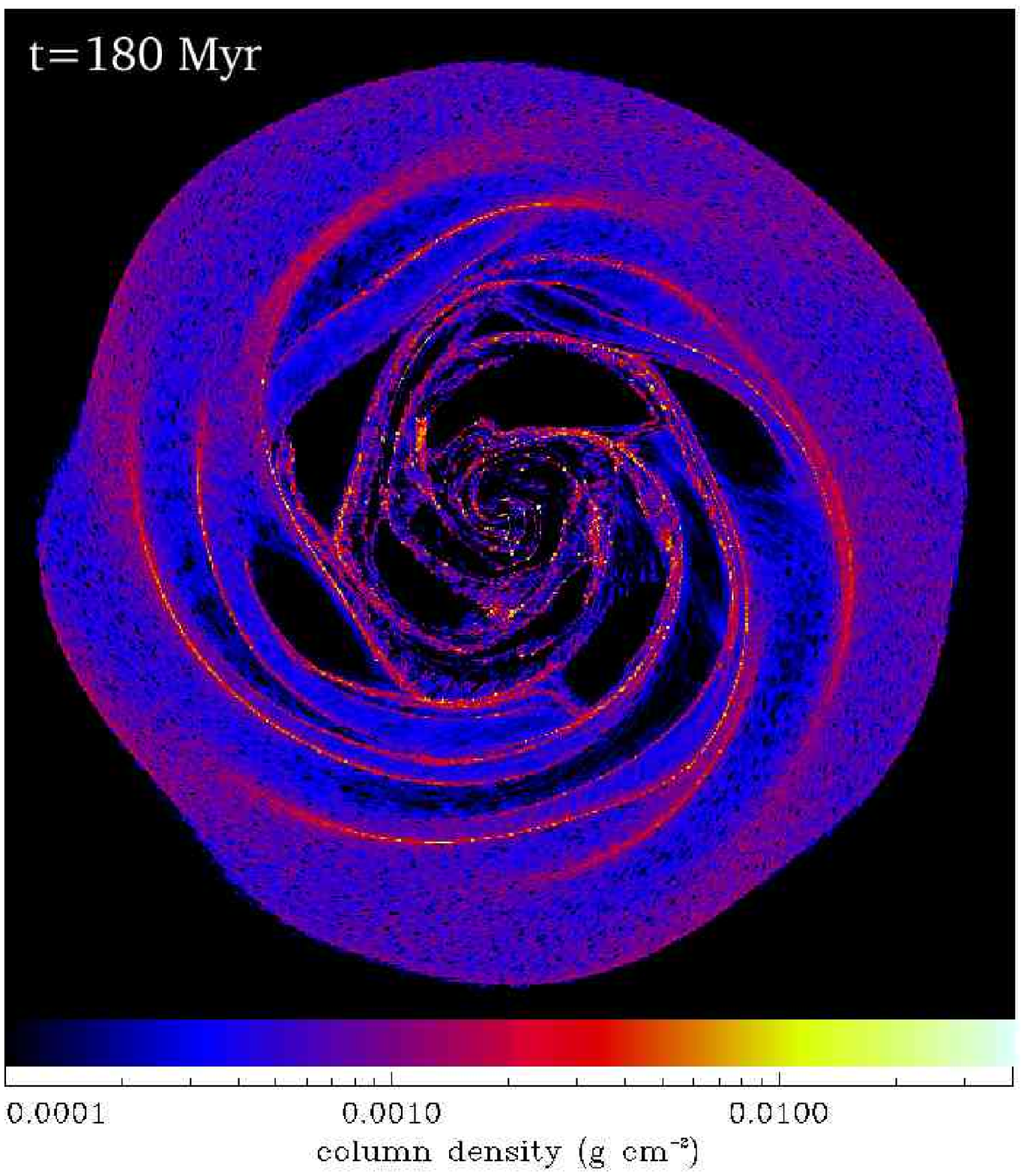}
\includegraphics[scale=0.33]{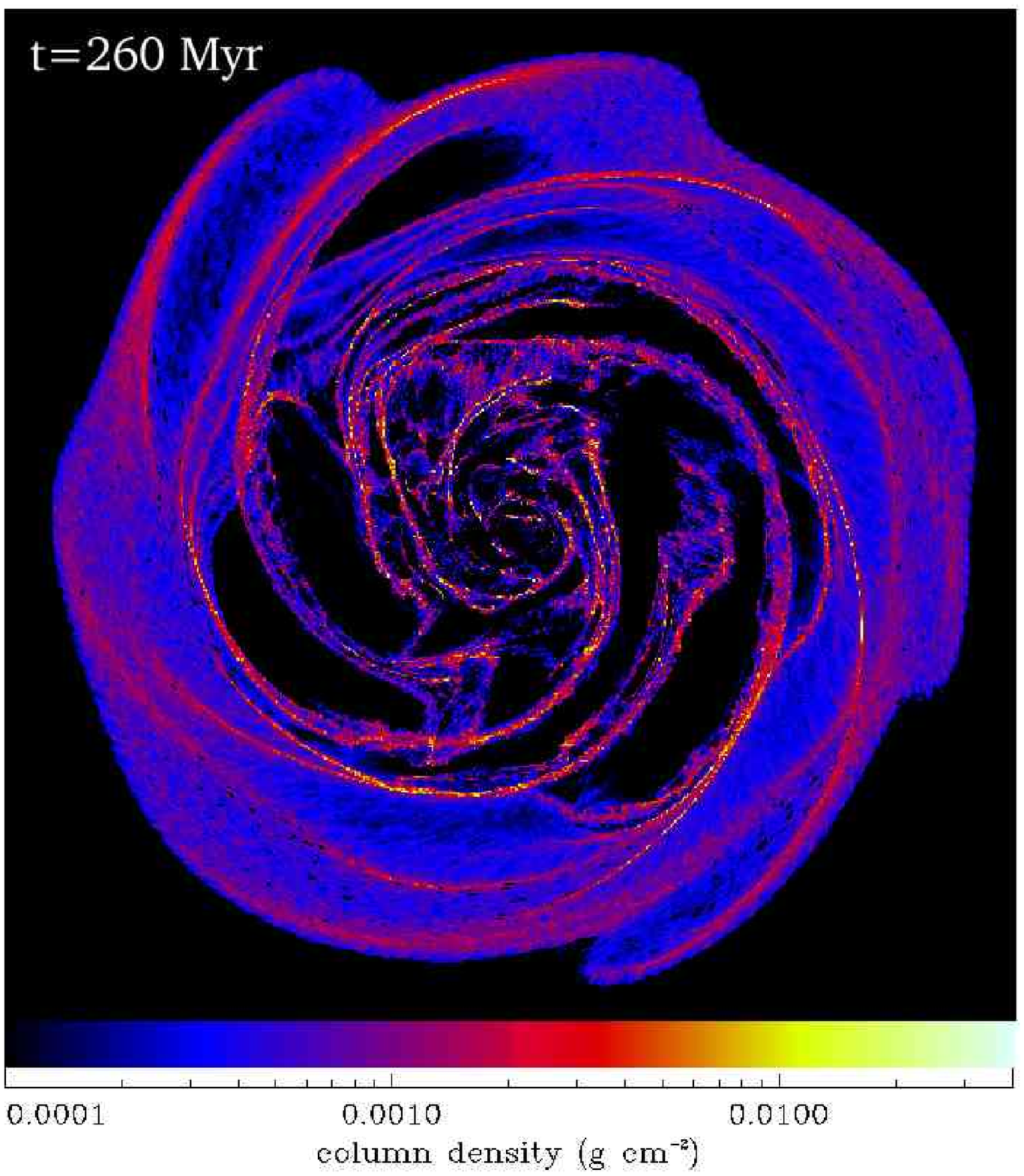}}
\vspace{2pt} 
\centerline{
\includegraphics[scale=0.33]{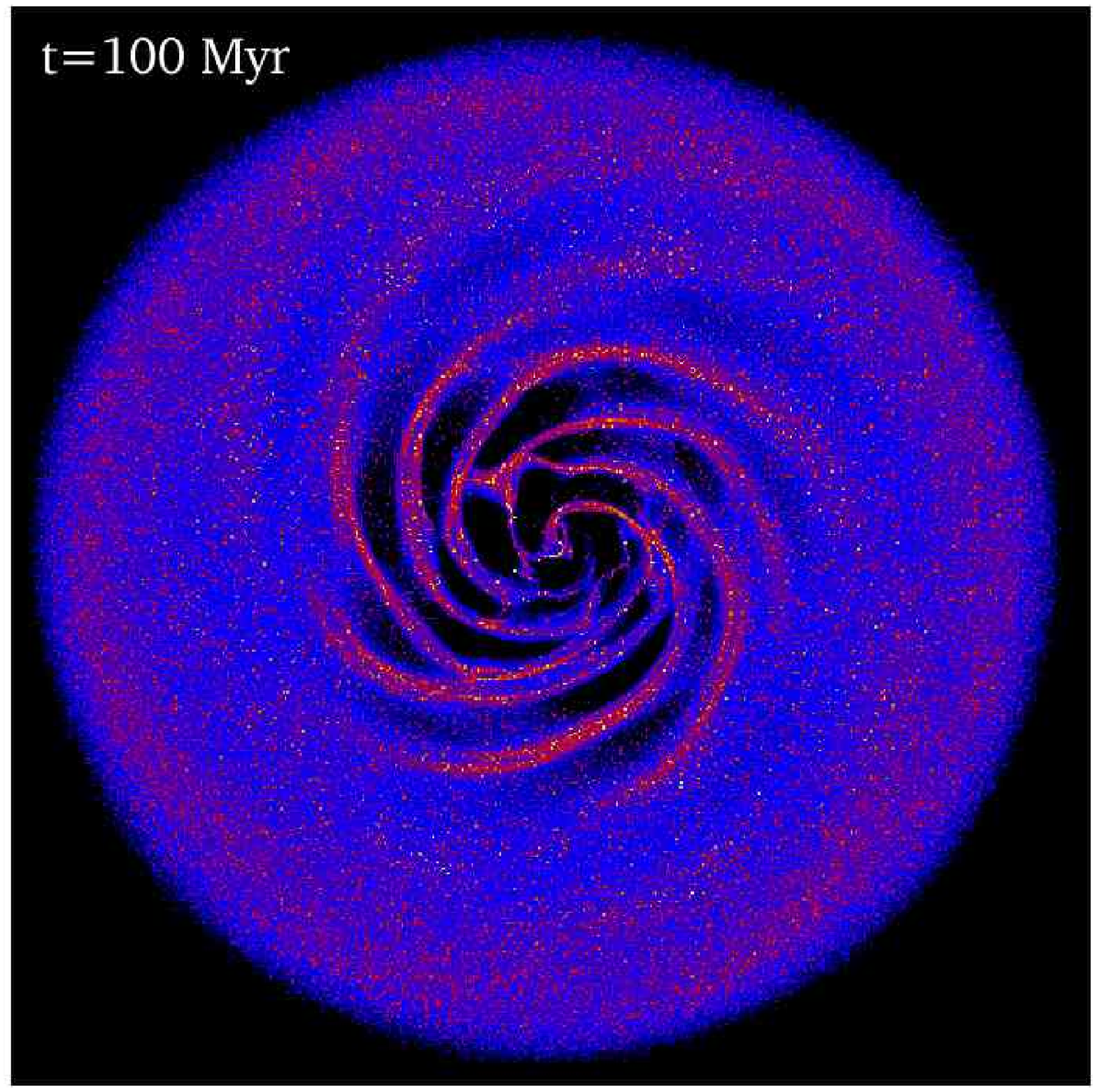}
\includegraphics[scale=0.33]{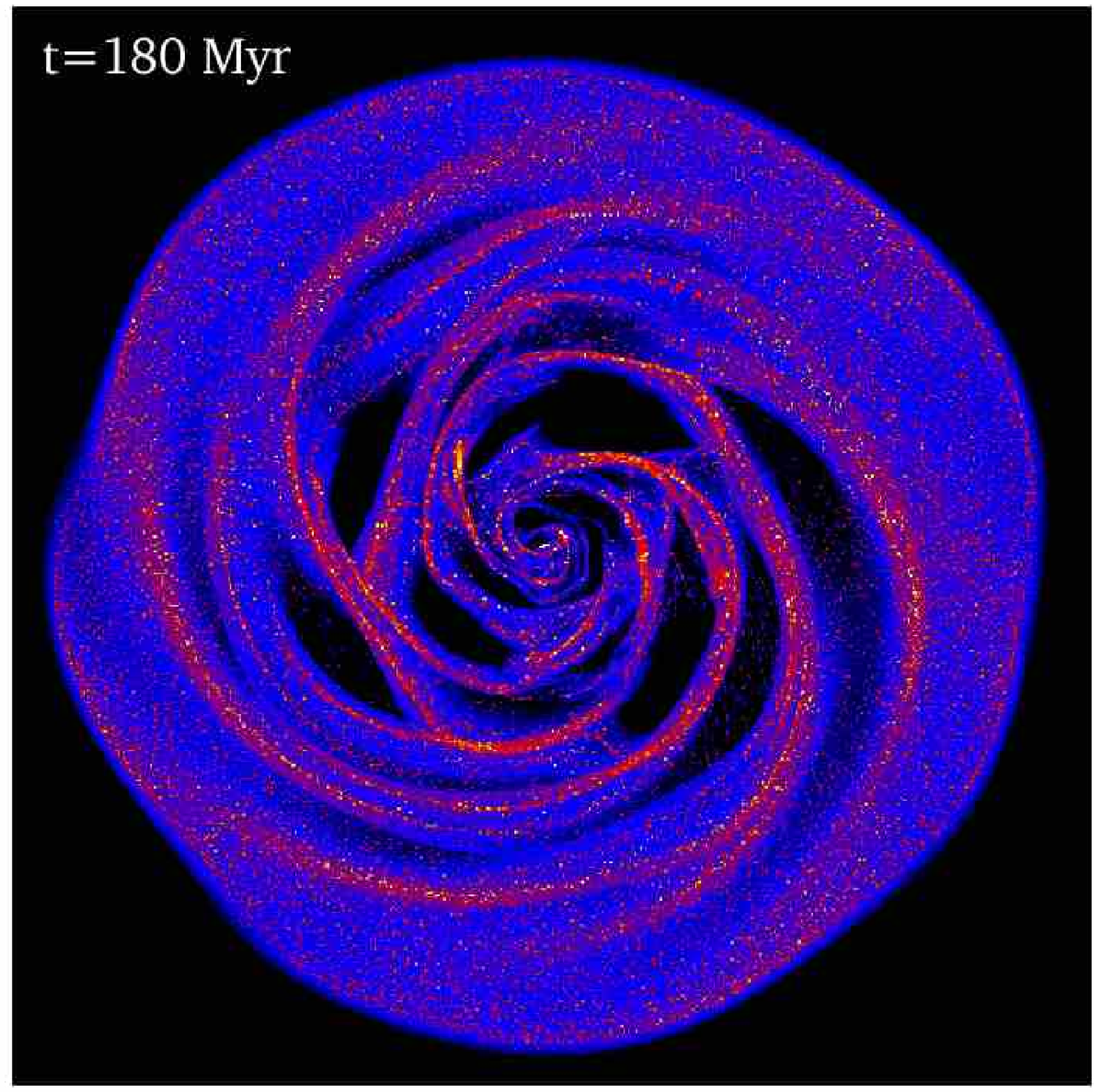} 
\includegraphics[scale=0.33]{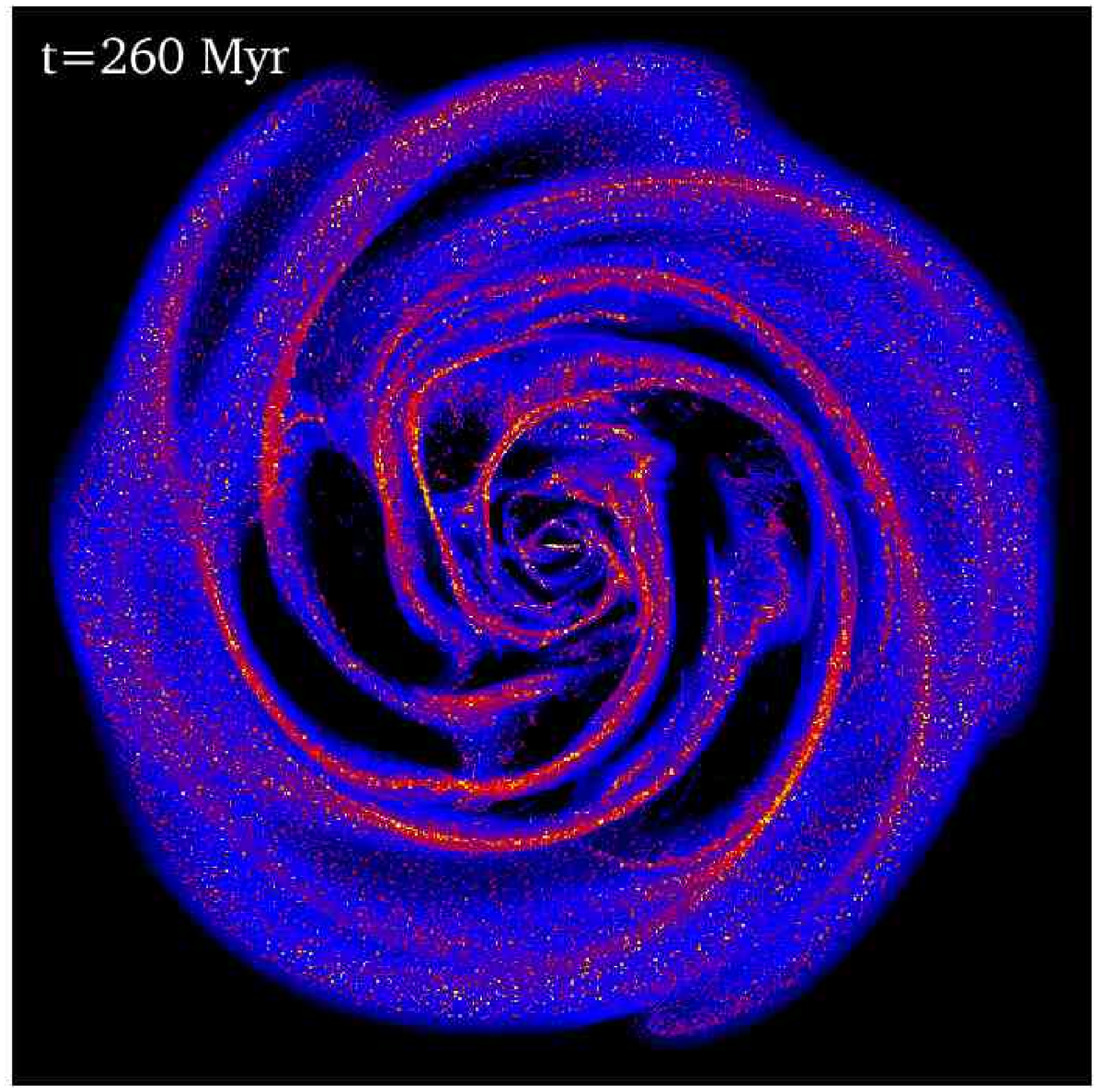}}
\vspace{5pt} 
\centerline{
\includegraphics[scale=0.33]{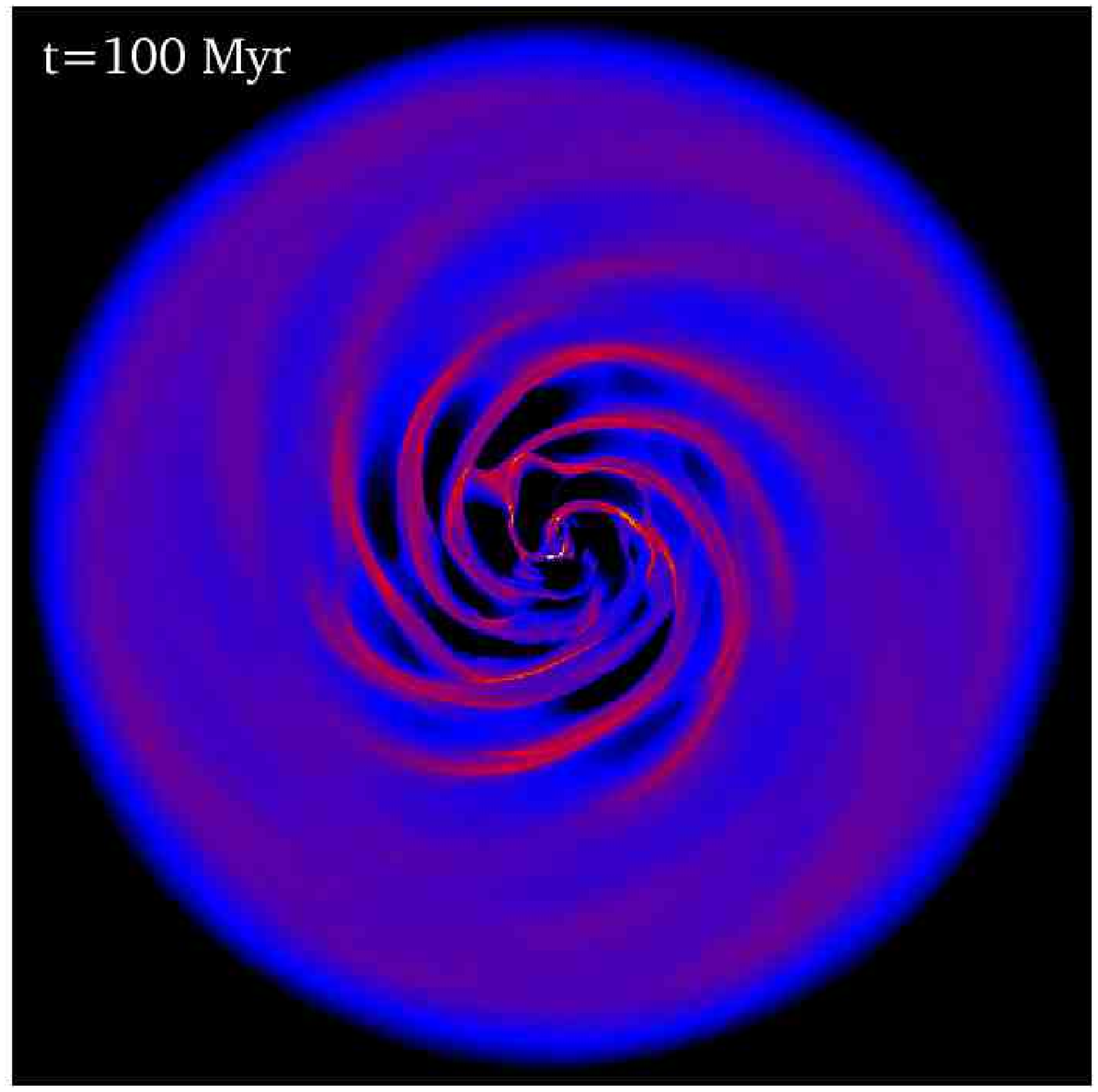} 
\includegraphics[scale=0.33]{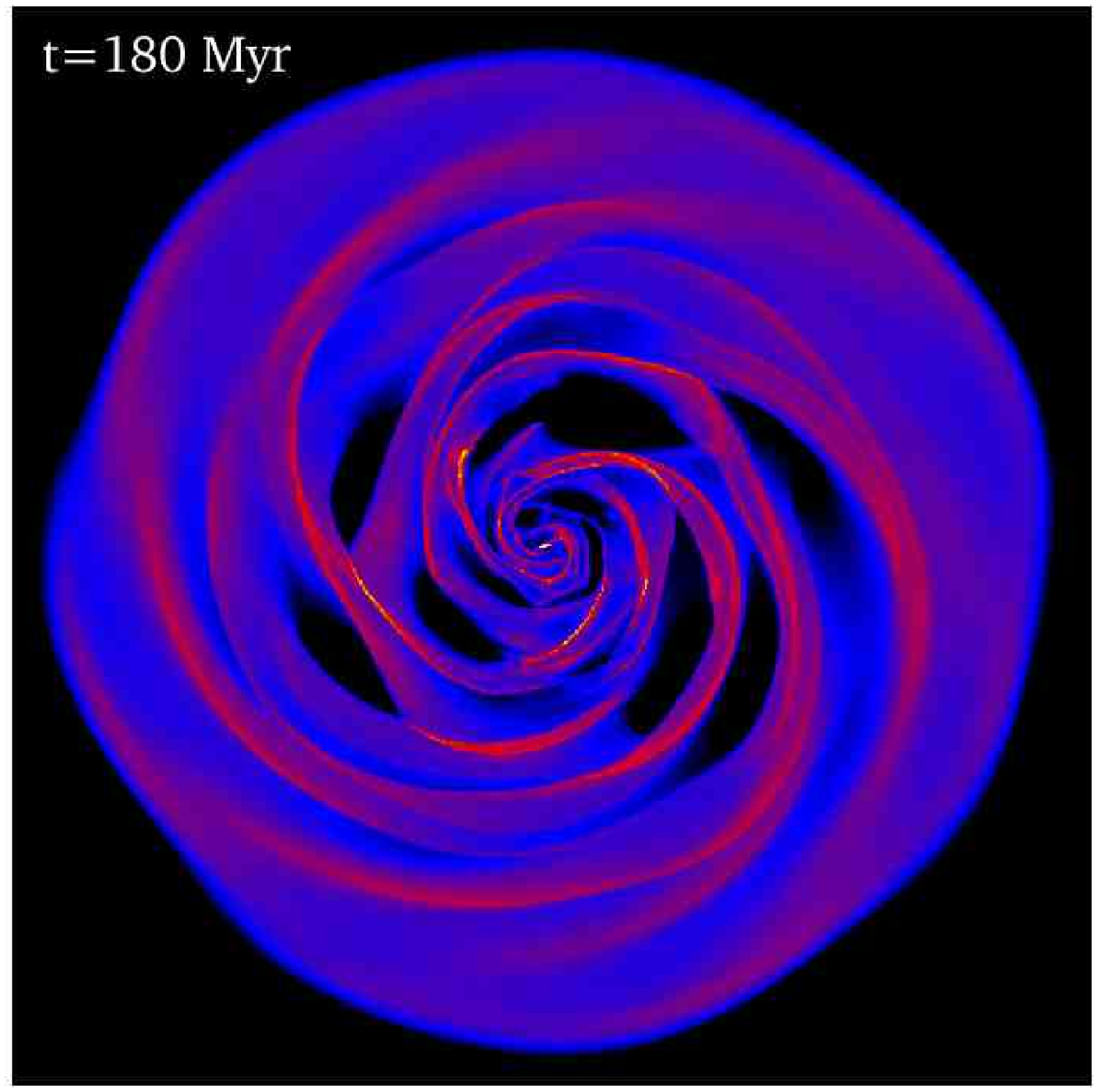}  
\includegraphics[scale=0.33]{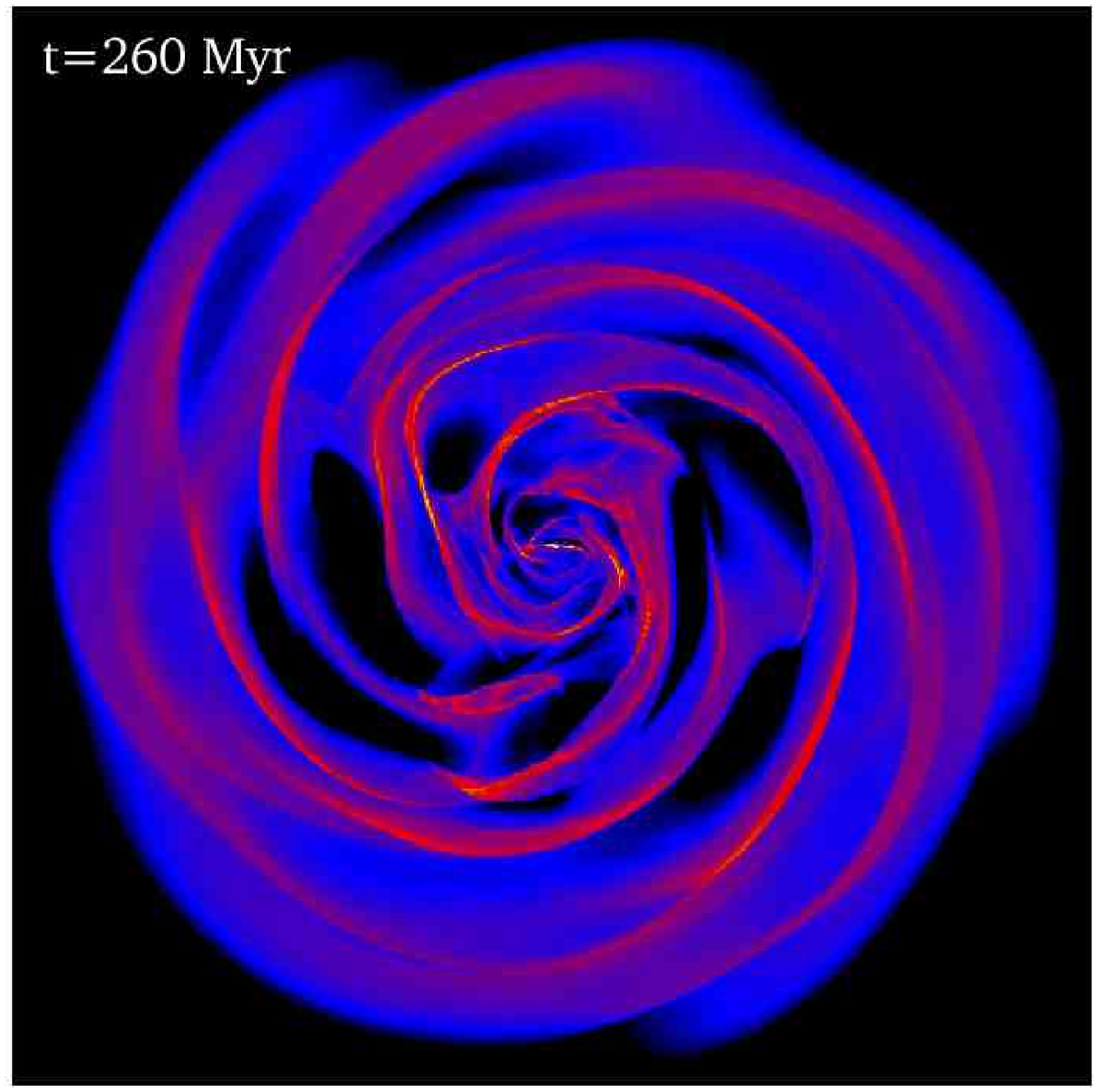}}
\caption{The time evolution of the spiral pattern is shown in these column density plots, which are inclusive of both cold and warm phases of gas. 
The top row shows Run A with predominantly cold gas, the middle with equal proportions of cold and warm gas (Run B) and for the bottom row, all the gas is $10^4$ K (Run C). 
Each plot shows a region of -11 kpc$<x<$11kpc and -11 kpc$<y<$11kpc, assuming a Cartesian axis centred at the origin, and gas is flowing anticlockwise round the disk. The main spiral arm features are analogous for the different thermal distributions, although there is much more structure on smaller scales when there is cold gas.}
\end{figure*}
The development of the spiral structure with time for the 3 different calculations is shown in Fig.~1. 
The perturbation to the potential, and hence the extent of the spiral arms, is at first restricted to the inner disc. The spiral arms spread to larger radii as the magnitude and extent of the spiral perturbation to the potential increase. 
The number of spiral arms is not constant and the dominant mode varies between $m=2$ and $m=8$ \citep{Clarke2006}.
The structure also changes as spiral arms collide with each other, giving the appearance of bifurcations of the arms into different branches. Enhanced star formation is predicted where high density regions develop as the arms collide \citep{Clarke2006}. 
Approximately half way through the N-body simulation, the spiral 
perturbation in the disk begins to decrease. Consequently the spiral structure begins to wind up after about 200 Myr in our simulations.

The large scale structure and spiral features (Fig.~1) are similar for the different thermal distributions. With cold gas present there are dense clumps in the shocks and higher density regions, whereas when there is only warm gas (lower panels), the spiral arms are smooth.
In some regions, the gas appears  to form `double shocks', much more noticeable for the cold gas, although still apparent for the simulation with warm gas and in \citet{Clarke2006}. These are due to new spiral arms forming, which attain similar rotational velocities to  older spiral arms. The structure towards the centre of the predominantly cold simulation becomes chaotic, due to repeated collisions of stronger shocks in this region.

A detailed section of the simulation with 50\% warm gas (Run B) is shown in Fig.~2. There is considerable structure in the cold gas in both the spiral arms and inter-arm regions. However, the agglomeration of gas into larger clumps in the shock is not  as apparent as when there is a standing density wave from a grand design potential \citep{Dobbs2008}. Instead, collisions between spiral arms are more important (e.g. Fig~2), and this is where the largest structures form in the gas (and thus where agglomeration of clumps is occurring).
\begin{figure*}
\centerline{
\includegraphics[scale=0.38]{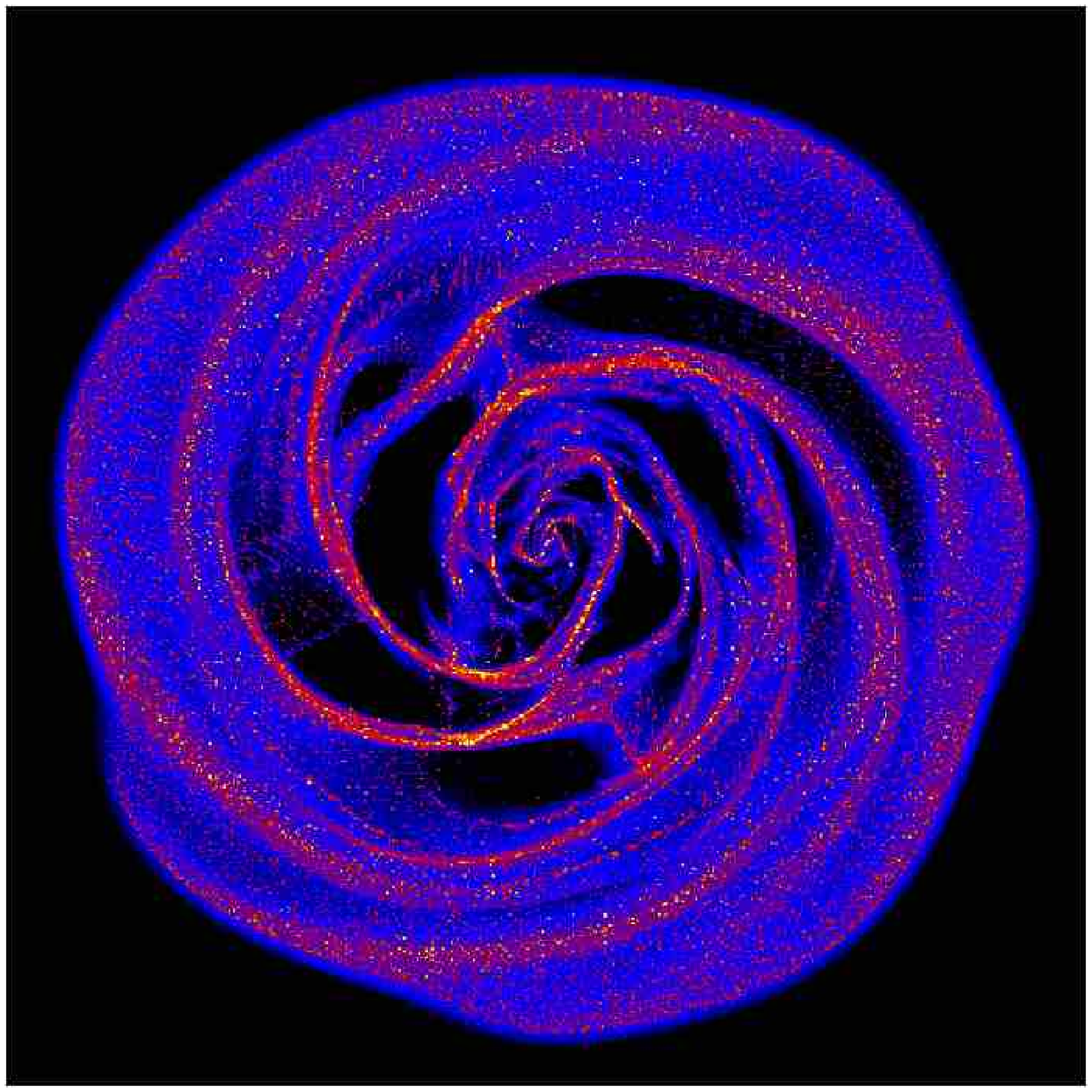} \hspace{10pt}
\includegraphics[scale=0.38]{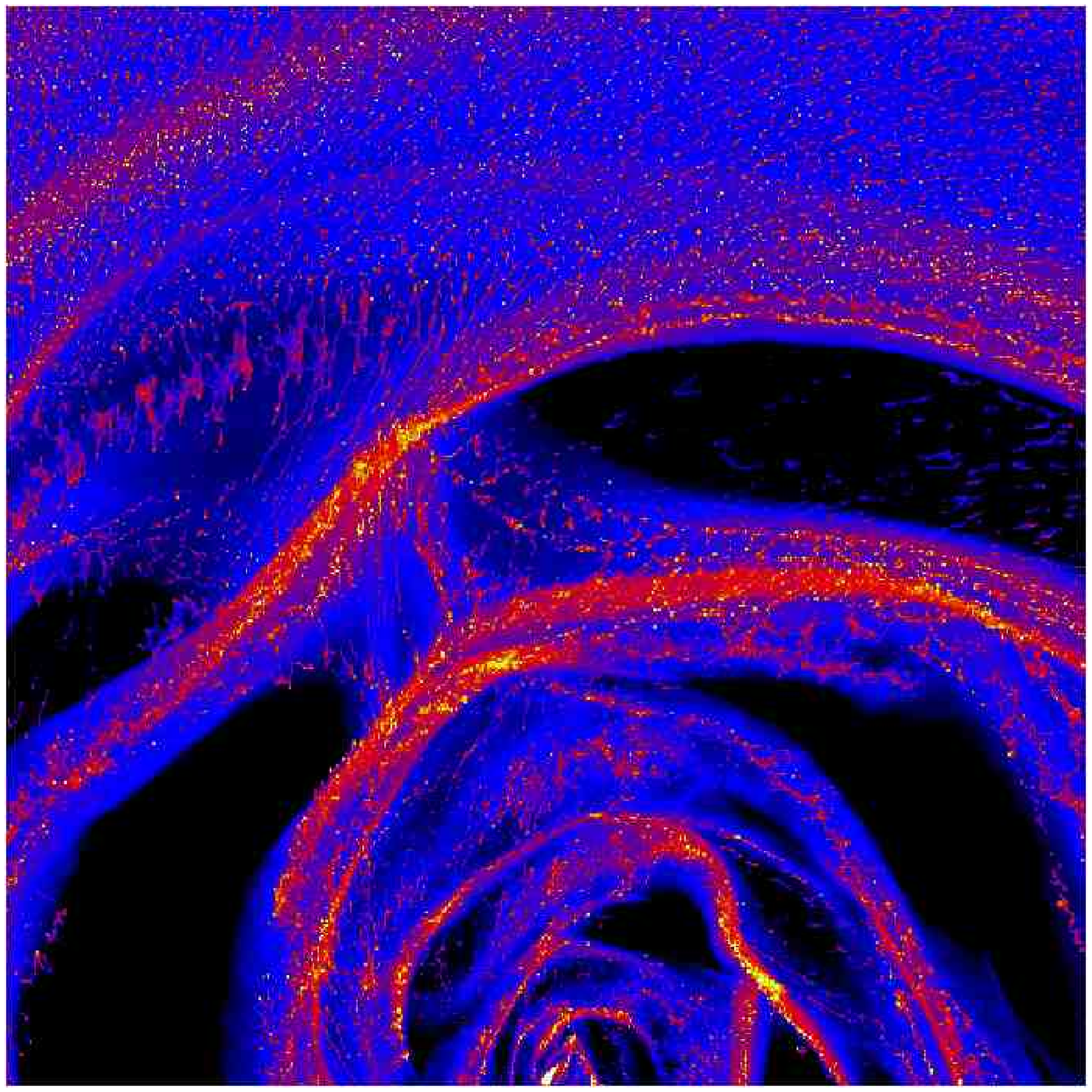}}
\vspace{5pt}
\centerline{
\includegraphics[scale=0.38]{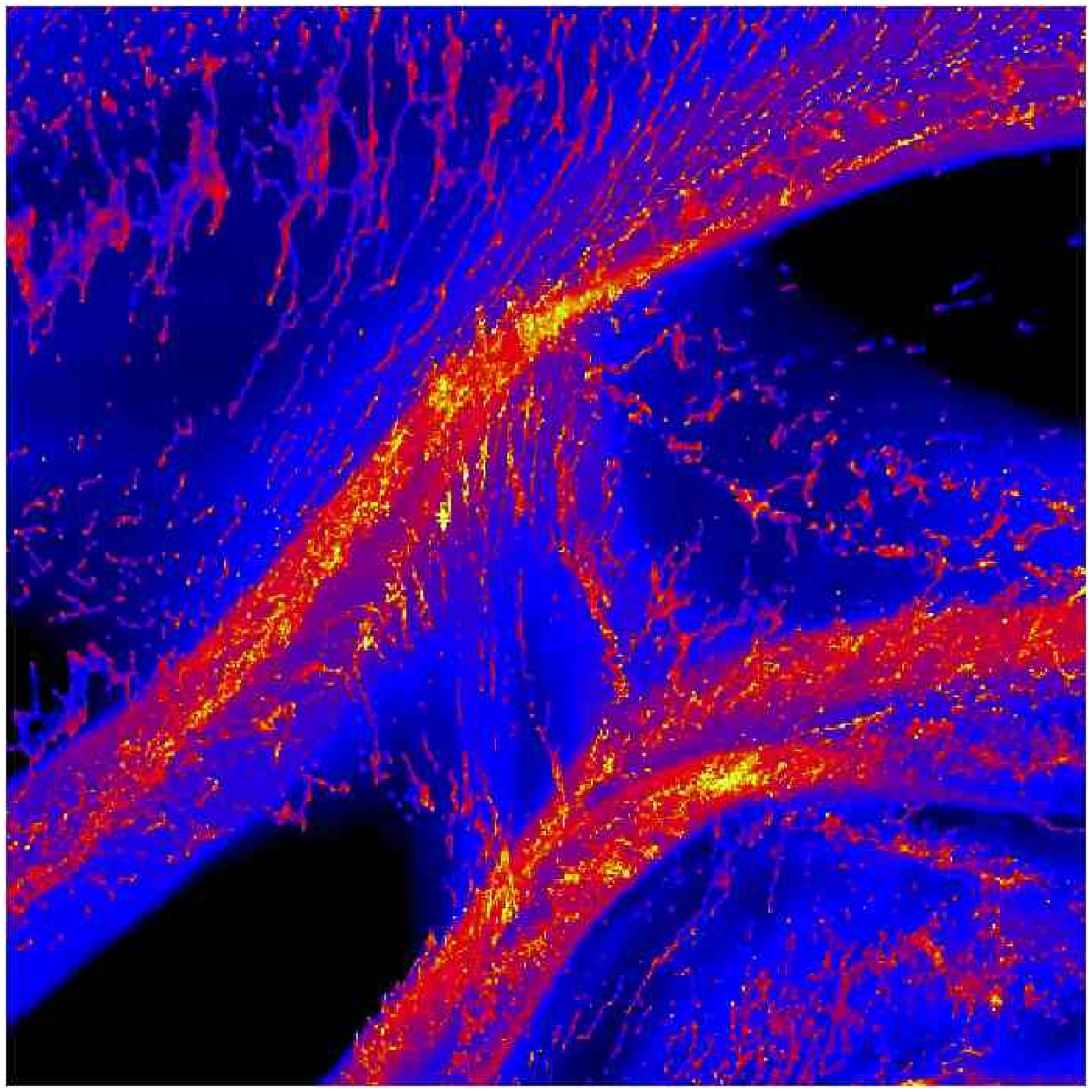} \hspace{10pt}
\includegraphics[scale=0.38]{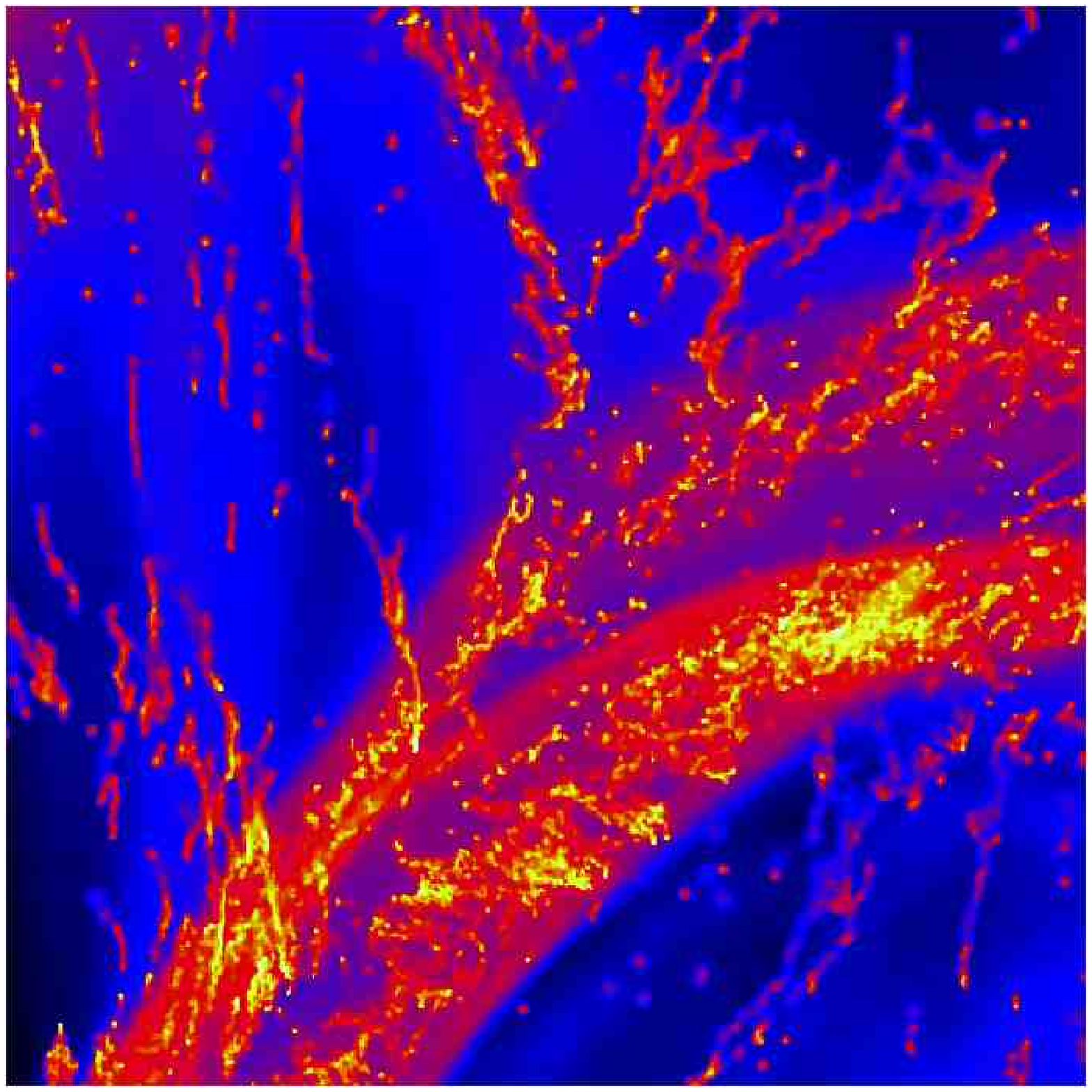}}
\caption{These images zoom in on a small region from the simulation with 50\% cold gas, Run B. The plots correspond to a time of 200 Myr, and show areas of size 
i) 22 kpc x 22 kpc, ii) 8 kpc x 8 kpc, iii) 4 kpc x 4 kpc and iv) 1.5 kpc x 1.5 kpc. The gas appears increasingly structured at smaller scales, becoming sheared into filamentary structures in the inner regions. The figure focuses on a bridge of gas between two spiral arms which corresponds with the remnants of a previous arm. Dense material has agglomerated either side of this bridge, where spiral arms have collided. The same scale is used as shown in Fig.~1. }
\end{figure*} 

\subsection{Gas dynamics of dynamically evolving galaxy}
The response of the gas to the active potential appears to be quite different to the behaviour of gas subject to a grand design potential.  We describe here the main differences between the two cases. 

For the active potential considered here, the gas essentially traces the potential minimum. This is in contrast to grand design galaxies, where the gas shocks as it passes the minimum but then leaves the spiral shock to continue to the next spiral arm.
Here gas enters the minima of the potential as they form, unlike the grand design case where the minima already exist. The gas then tends to accumulate in the potential minima rather than climbing out of the potential, retaining spiral structure even as the potential minimum dissolves. 
\begin{figure*}
\centerline{
\includegraphics[width=80mm]{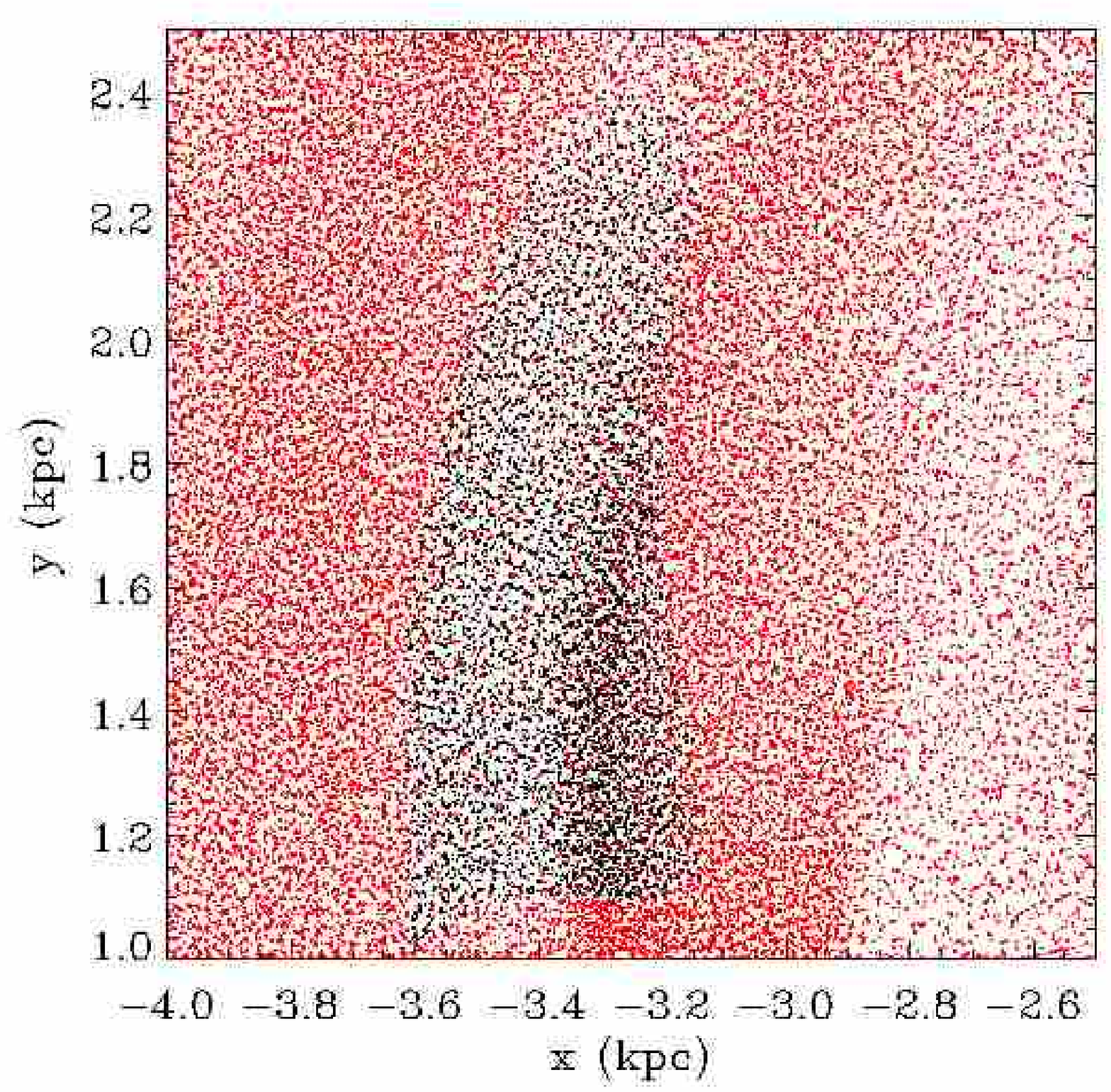}
\includegraphics[width=80mm]{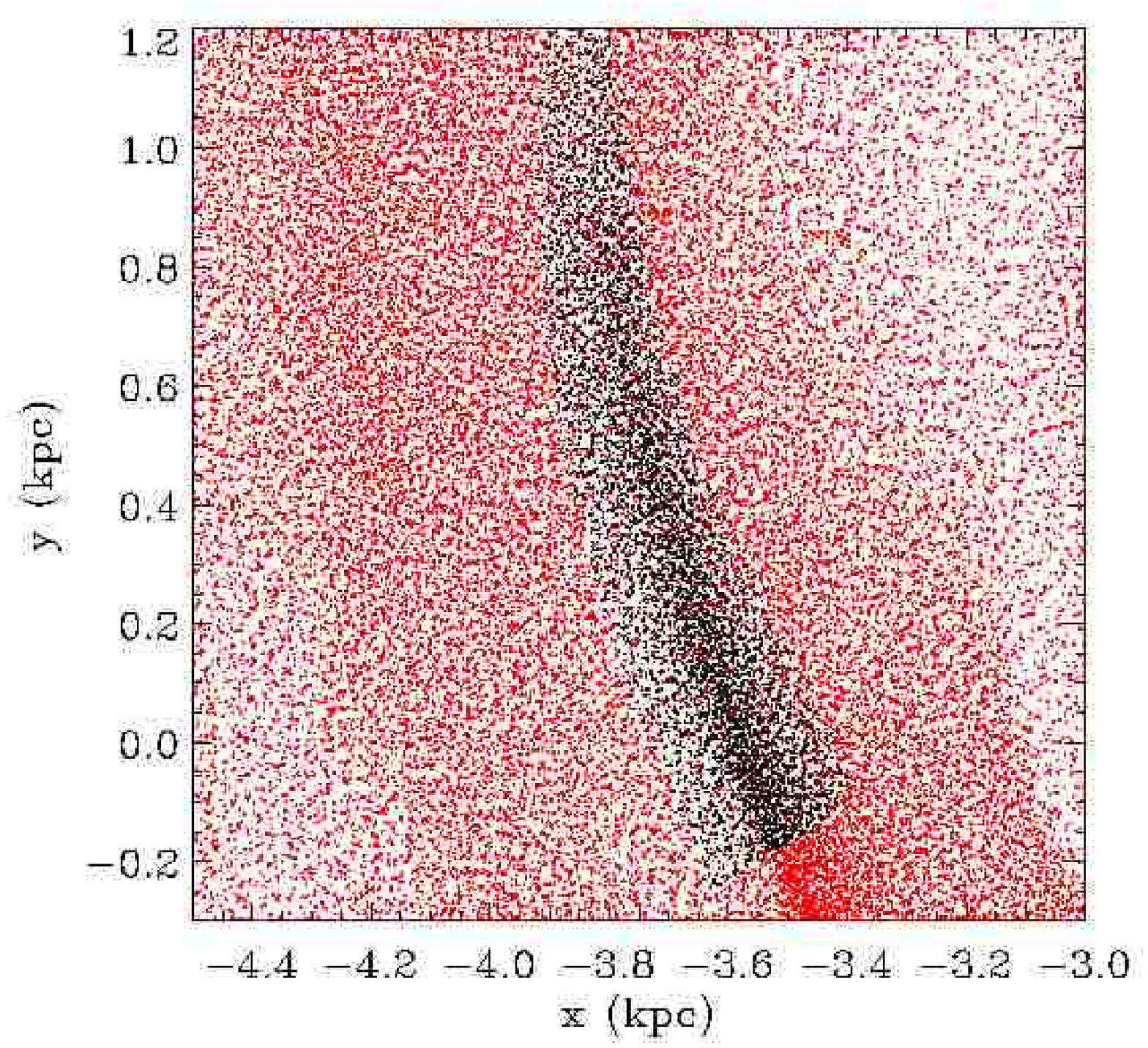}}
\centerline{
\includegraphics[width=80mm]{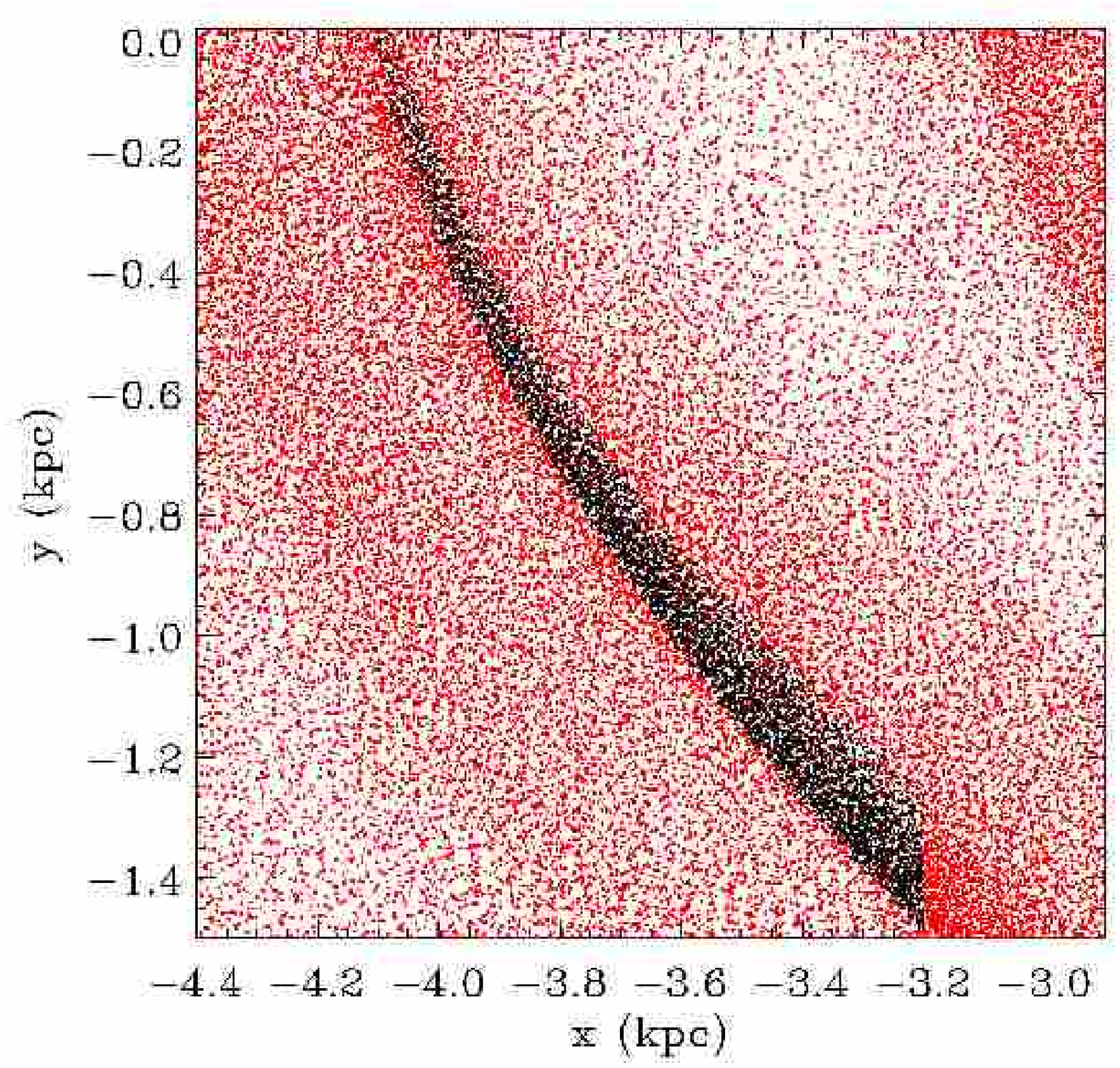}
\includegraphics[width=80mm]{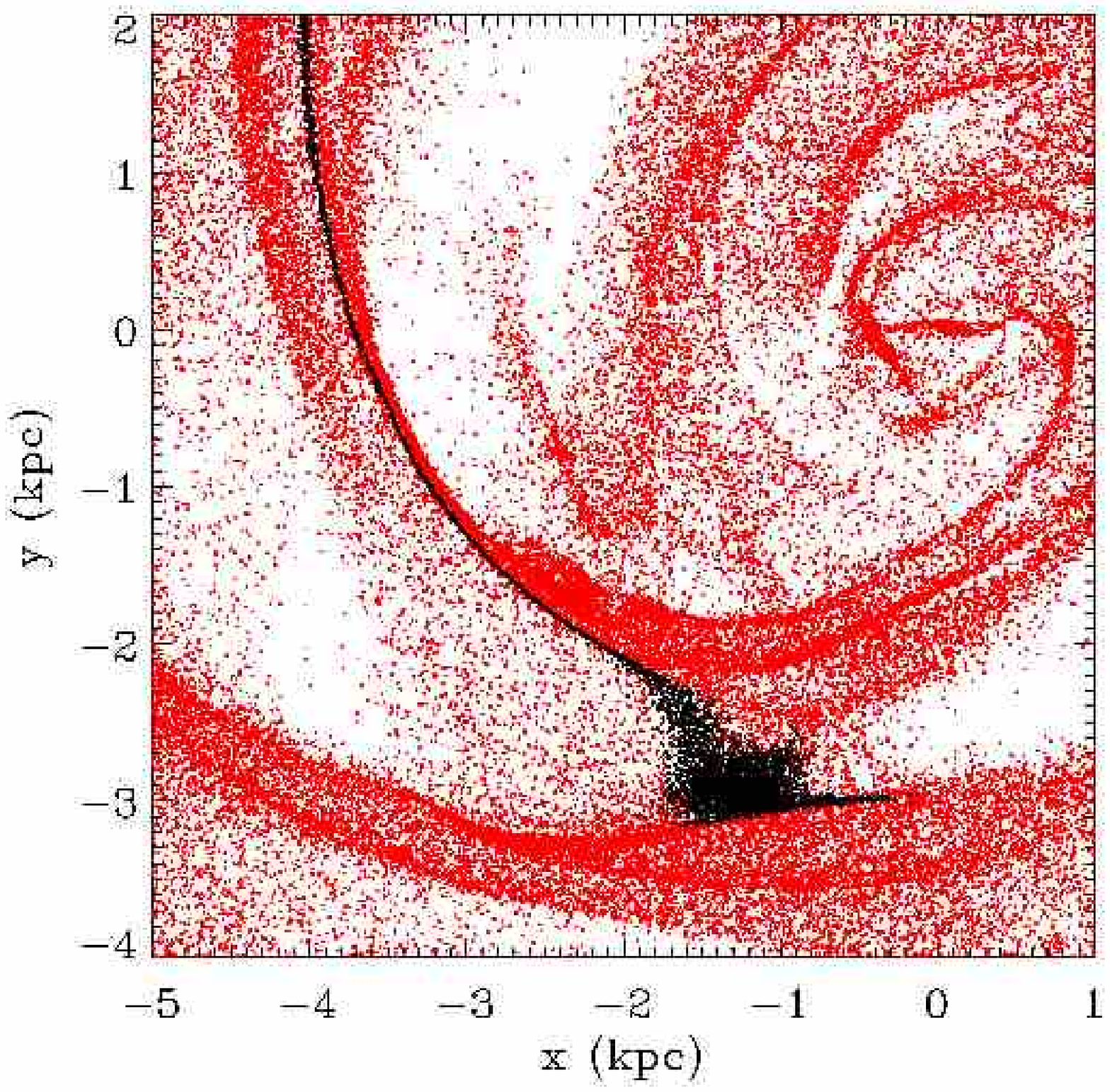}}
\caption{The particles are shown for a subsection of the disc at times of 70 (top left), 76 (top right), 82 (bottom left) and 195 (bottom right) Myr for the simulation with warm gas (Run~C). Particles indicated in black were selected from the spiral arm displayed in the bottom left panel, and are shown at earlier and later times in the other panels. The first three panels show the accumulation of gas into a spiral arm. The spiral arm structure of the gas persists as the gas encounters multiple spiral arm passages, as indicated by the fourth panel. At the latest time, only 1 in 10 of the total number of particles are plotted.}
\end{figure*}

In Fig.~3, we show the typical evolution of gas as it enters a potential minimum. The figure highlights a selection of particles at 4 different time frames from the simulation with warm gas. The particles were selected as those in the section of spiral arm shown in the third frame, with their locations at earlier times shown in the top two panels. 
As a minimum develops in the underlying potential, a shock occurs as gas falls into the minimum and produces a spiral arm.
The top two panels of Fig.~3 indicate that particles enter the minimum from both sides of the developing spiral arm. The spiral arm in the bottom left panel is much sharper on the upstream side where the gas generally flows into the spiral arm \citep{Bonnell2006}. 
The fourth panel in Fig.~3 (bottom right) shows the particles at a much later time, by which time they have formed a completely different spiral arm.

Fig.~4 shows the densest gas from 2 of the simulations overplotted on the underlying potential. In the top panel, the gas is taken from the simulation with predominantly cold gas (Run A) and for the other, the gas is warm (Run C). Only the non-axisymmetric component of the potential is shown, calculated at each point by subtracting the mean value of the potential over the corresponding cylindrical radius.   
In both cases the high density gas strongly coincides with the potential minimum. Since gas effectively falls into the spiral potential, this is not surprising. 
The coincidence between the spiral arms and the potential minima is in contrast with grand design models where the spiral shocks tend to be offset from the potential minimum. \citet{Roberts1969} found that spiral shocks occur on the leading side of the potential minimum inside co-rotation for quasi-stationary solutions with $m=2$. However the location varies depending on sound speed, pattern speed, number of spiral arms and magnetic field strength, as shown in recent simulations 
\citep*{Slyz2003,Gittins2004,DP2007,Wada2007}. For the active potential applied here, a quasi-stationary state of course does not evolve, since the spiral potential changes with time, and we would not expect an offset between the spiral shock and star formation. 

There are nevertheless some sections of spiral arms that appear to cross between two minima.The stellar spiral arms associated with the N-body simulation are transient, so the minimum will eventually disappear. However the gaseous spiral arms still retain their spiral structure, even though the minimum dissolves. Eventually the gas reaches another minimum and joins a new spiral arm. The final panel in Fig.~3 shows a different section of spiral arm at a much later time frame. The selected particles have already passed 3 spiral arms at this point.  
Gas particles highlighted with $y>-2$ kpc are largely coincident with a potential minimum, and also corresponds to one of the dense arms of gas in Fig.~4.  On the other hand, gas particles with $x>-2$ kpc  lie between two minima or are forming the new spiral arm seen in the lower part of the figure. In Fig.~4, spiral arms lying between the potential minima are most evident when cold gas is present and the densities of the gas are higher.
A further consequence of gas retaining the spiral structure imposed by the potential is that at later times in the simulations (e.g. 3rd time frame, Fig.~1), most of the gas is in the spiral arms. Regions between the arms are relatively empty, and since little gas is entering the potential, the spiral arms correspond merely to dense regions of gas rather than actual spiral shocks.

In addition to the density, the velocity of the gas which is in the spiral arms largely follows that of the stellar component of the disc, with different pattern speeds emerging for different sections of spiral arm. The rotational velocities of the gas thus indicate that there is no fixed pattern speed or co-rotation radius for the spiral perturbation.
\begin{figure}
\centerline{
\includegraphics[width=64mm]{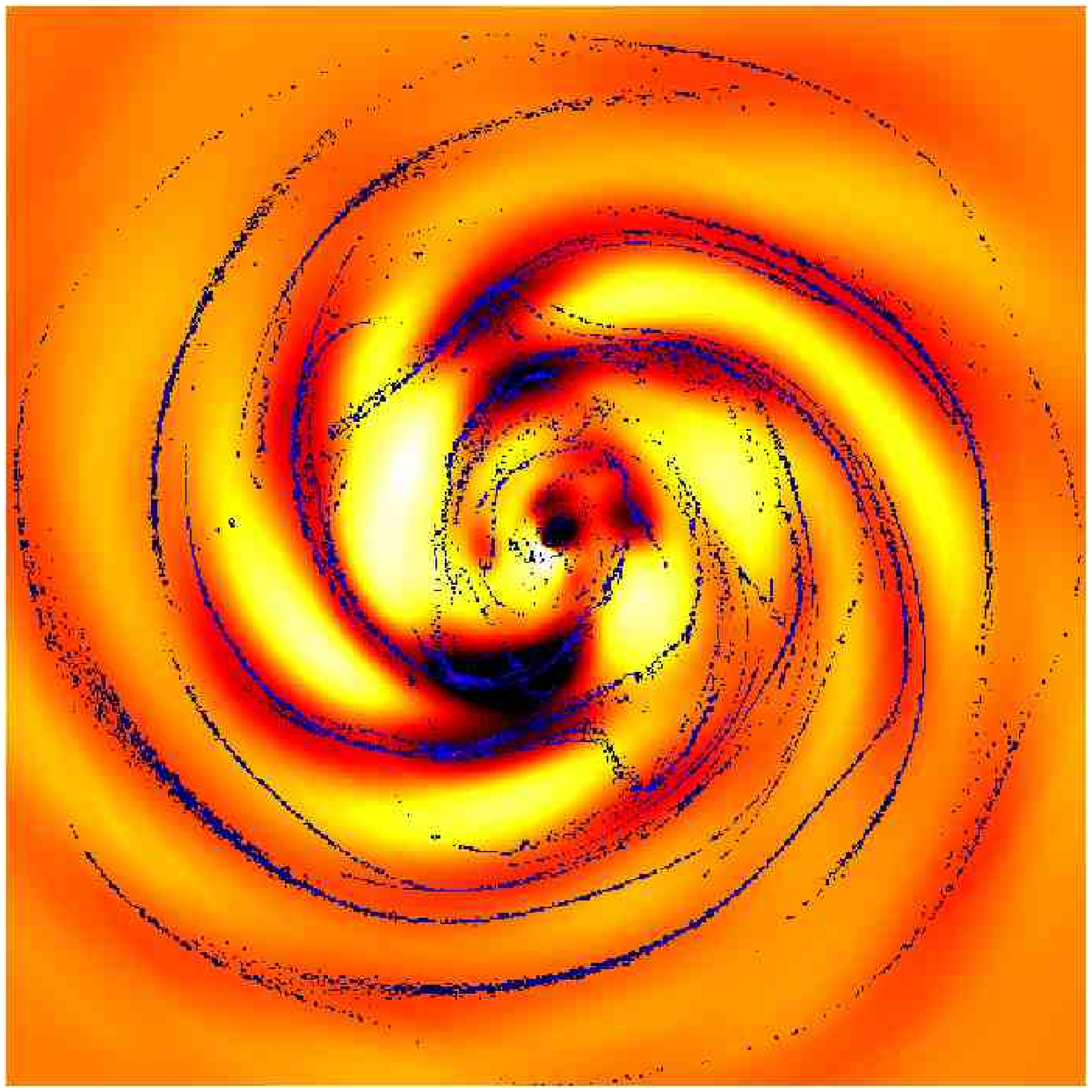}}
\vspace{5pt} 
\centerline{
\includegraphics[width=64mm]{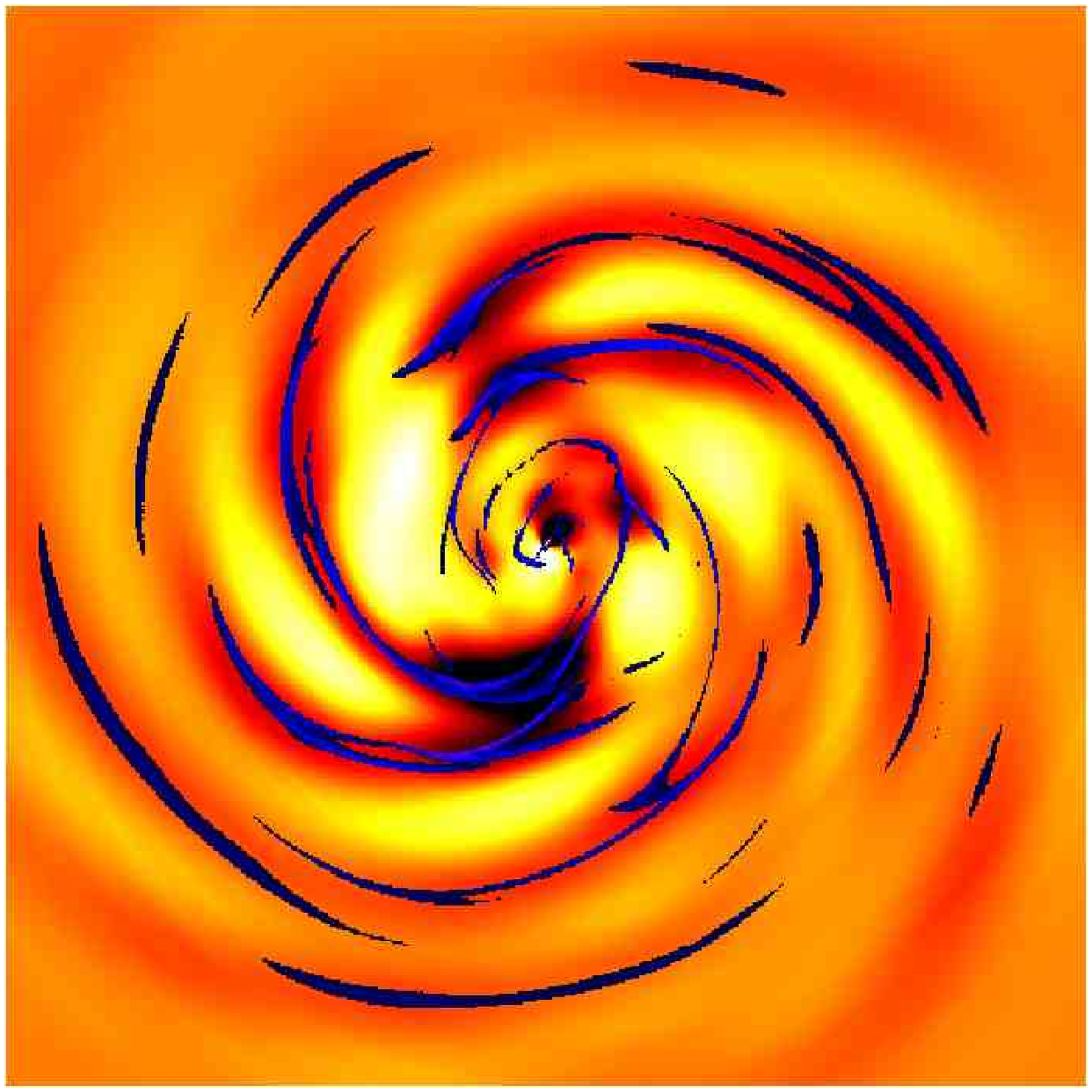}}
\caption{The most dense gas from simulations with predominantly cold (Run A, top) and warm (Run C, lower) is overplotted on the potential from the N-body simulation. Only the spiral component of the potential is indicated, and black/dark red indicate regions where the potential is lowest, yellow/white indicates the potential is highest. The gas is shown in blue. Generally the densest gas corresponds with the minima of the potential. The regions shown are 20 kpc x 20 kpc and the potential rotates in the anti-clockwise direction. The corresponding time is 200 Myr.}
\end{figure}

\subsection{Velocity dispersion}  
Fig.~5 shows the velocity dispersion for the 2 phase simulation with equal components of warm and cold gas (Run B) as a function of azimuth. The average velocity dispersion is calculated over a ring of width 200 pc and azimuthally divided into 100 segments. The ring is centred at 2 different radii, 4 kpc and 8 kpc. Both show peaks where the gas passes through spiral shocks, but whereas at radii of 4 kpc, the velocity dispersion reaches $>10$ km s$^{-1}$, at 8 kpc the velocity dispersion barely exceeds 3 km s$^{-1}$.  
All the gas (warm and cold) is used to determine the velocity dispersion in Fig.~5. The variation in the velocity dispersion is dominated by the cold gas.
 
 The increase in the velocity dispersion occurs as clumpy gas passes through a spiral shock \citep{Dobbs2007a} and the magnitude of the increase is dependent on the Mach number of the shock. At smaller radii, the potential is much stronger, so the Mach number and therefore velocity dispersion are higher.
Likewise, for a given radius, the velocity dispersion in the spiral arms increases with time whilst the strength of the perturbation to the potential increases. However the velocity dispersion starts decreasing by 250 Myr, since little gas is entering the spiral arms and shocks are no longer occurring. 
\begin{figure}
\centerline{
\includegraphics[width=90mm]{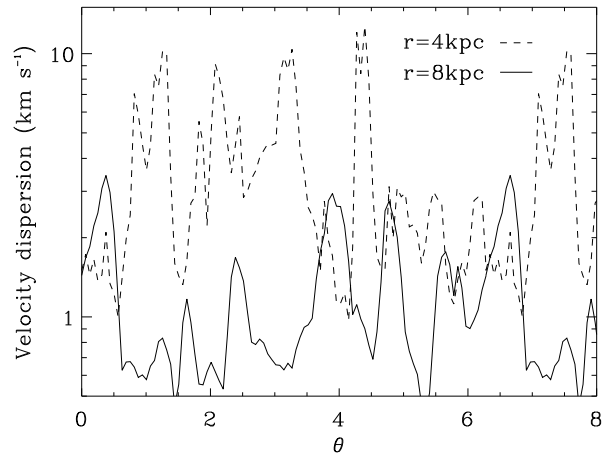}}
\caption{The velocity dispersion is shown against azimuth after 200 Myr for Run B (50\% cold gas). The angle $\theta$ is measured anticlockwise, in the direction of the flow, at radii of 4 and 8 kpc.}
\end{figure}

\subsection{Spurs and branches}
In simulations of grand design galaxies, regular spurs are found to extend from the arms into the inter-arm regions. These spurs are due to the shearing of clumps of gas leaving the spiral arms. They can occur if there is cold gas ($<1000$ K) without self gravity \citep{Dobbs2006}, or evolve from gravitational instabilities in warm gas \citep{Kim2006}. These features are not apparent in the simulations with the active potential. As mentioned in Section~3.2, gas tends to stay in the spiral arms and retain spiral structure, even as the minimum dissolves. Thus without gas leaving the spiral arms, gas cannot form spurs. From a survey of feathering in spiral galaxies \citep{LaVigne2006}, 20\% of galaxies are found to have periodic feathers. Generally galaxies with complex or flocculent structures are less likely to contain clearly defined feathers, although those with clearly defined dust lanes are more likely to contain such features.

Instead, when a minimum in the potential disappears, the gas retains the shape of the whole arm. Thus much larger branches occur in these simulations, located between the main spiral arms. A clear example is shown in Fig.~2b. These features are not present in simulations with 
time independent potentials, except where resonances occur \citep{Patsis1997,Chak2003}. As the potential minimum decays, the velocity of the spiral arm adjusts to that of circular velocities, and the shape of the spiral arm changes by differential rotation.
Large branches can be seen extending from the main spiral arms in many galaxies such as M101, M74 and M33. 
The pitch angle and length of these features tend to vary much more than would be expected from spurs which originate from instabilities in the spiral arms.
\begin{figure*}
\centerline{
\includegraphics[scale=0.4]{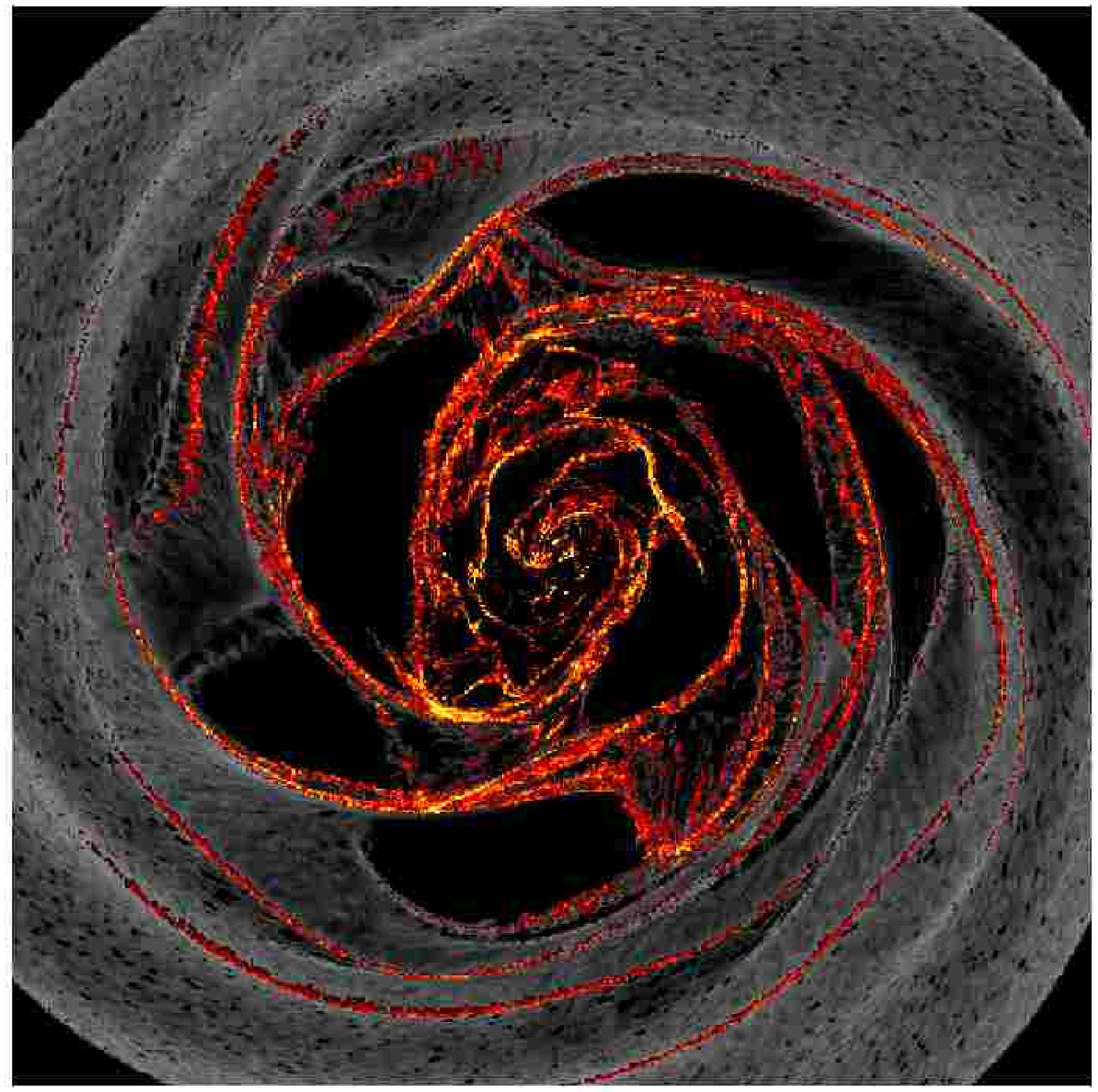} \hspace{5pt}
\includegraphics[scale=0.4]{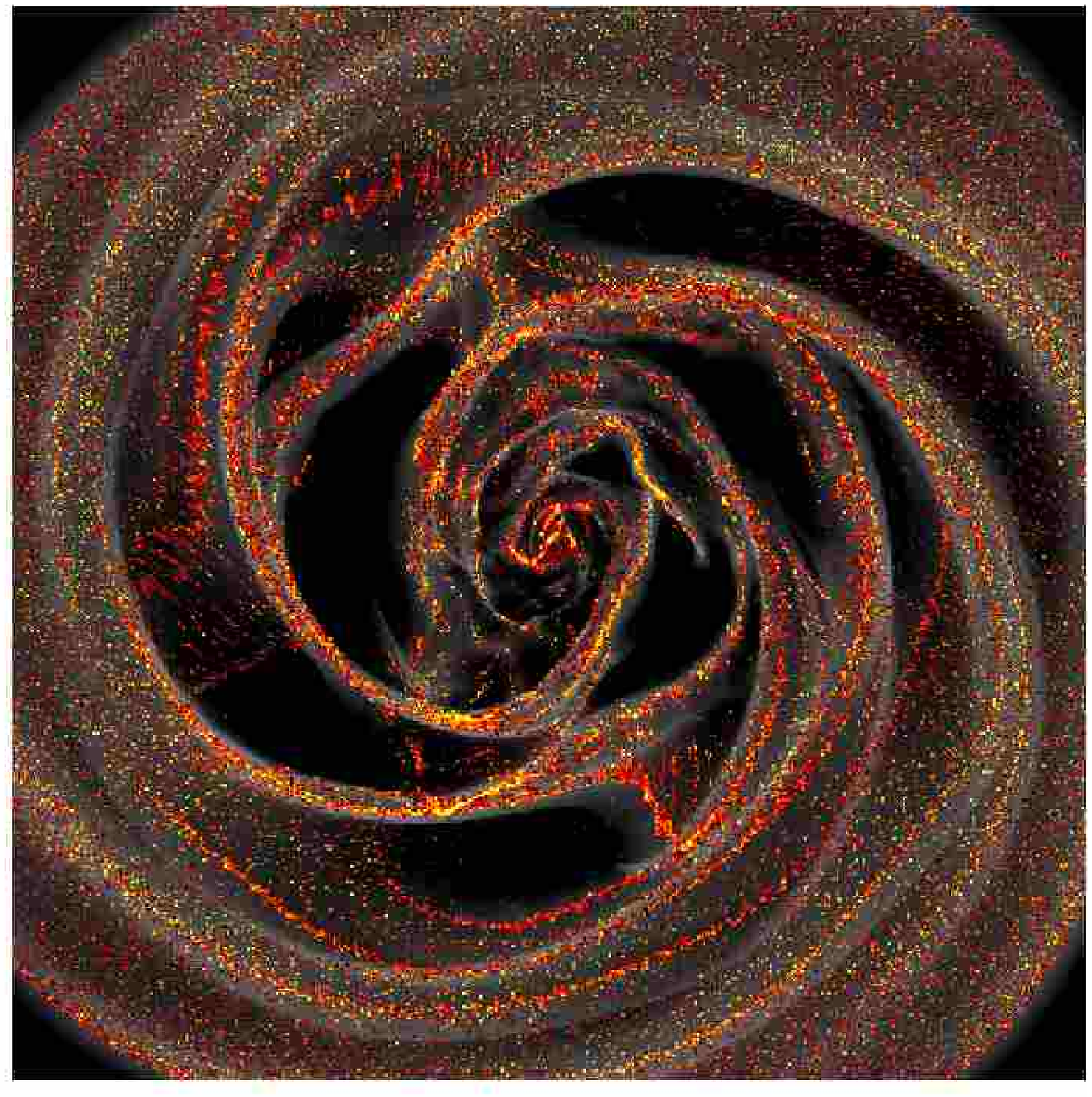}}
\centerline{
\includegraphics[scale=0.4]{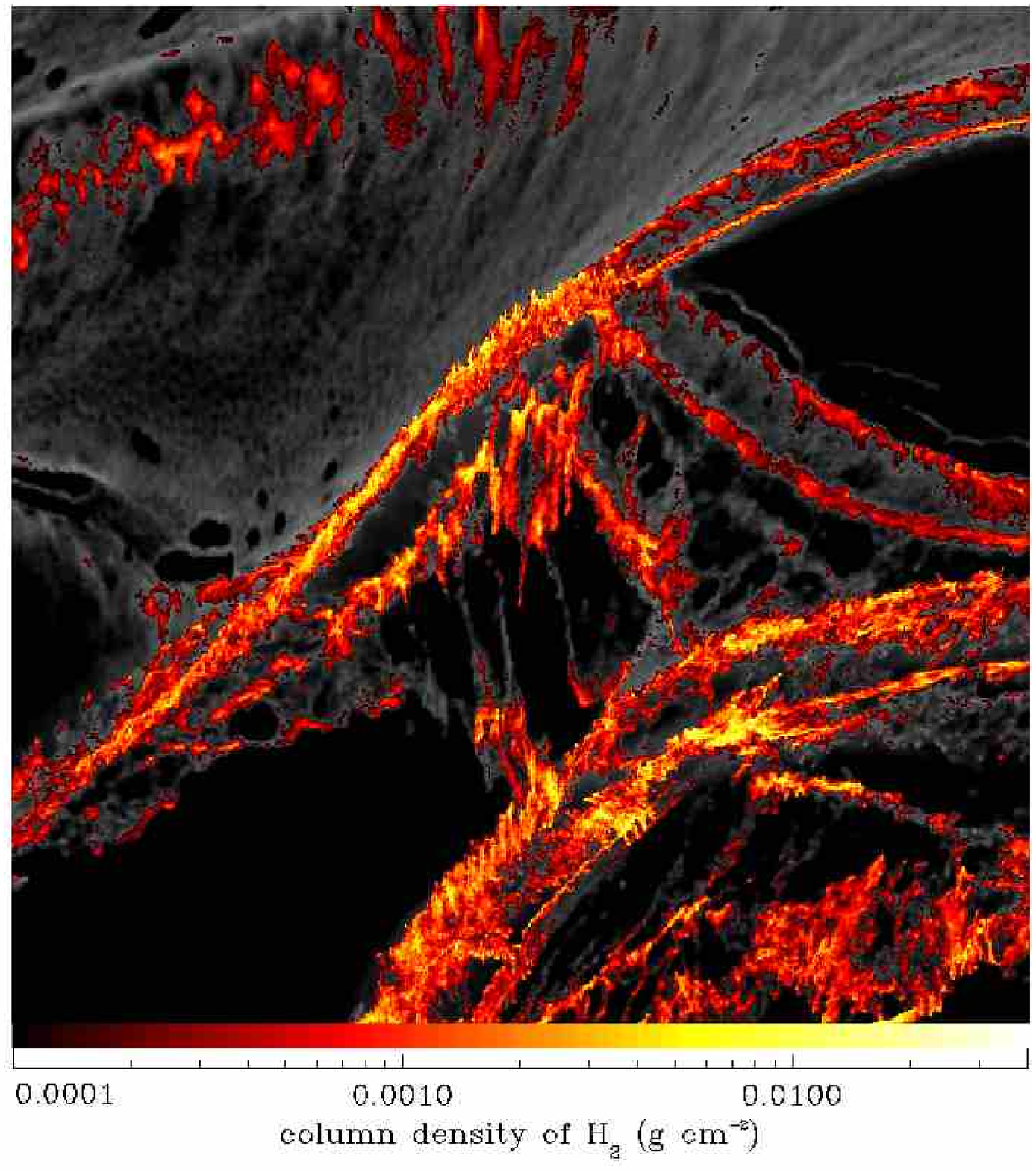} \hspace{5pt}
\includegraphics[scale=0.4]{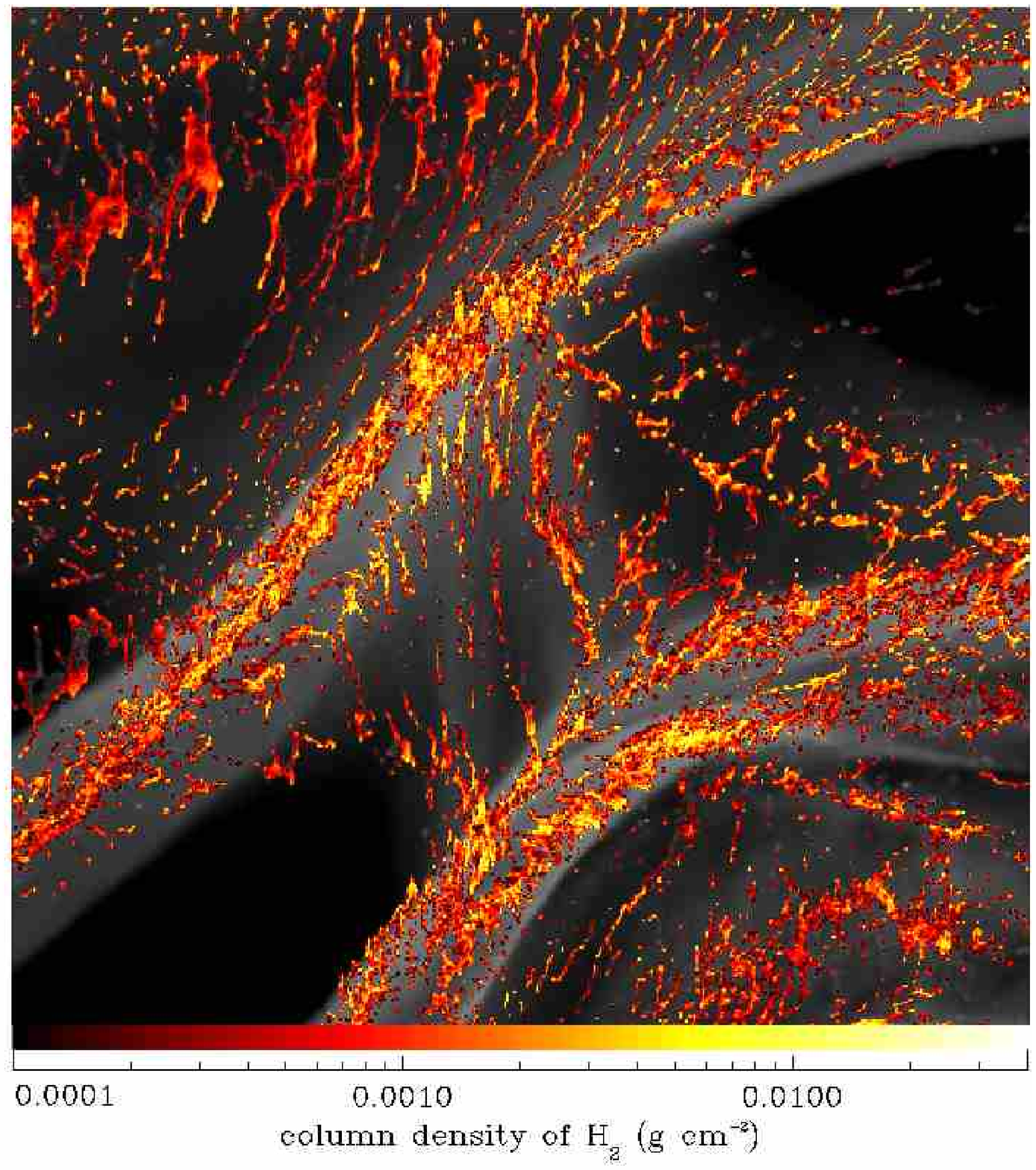}}
\caption{These images display the column density of molecular hydrogen (red) overplotted on the total column density of HI, from Runs A and B. The left panels show Run A,  with 90\% cold gas, and the right  Run B, with 50\% cold gas. The top panels show regions of 16 kpc x 16 kpc, and the bottom panels show an area of 4 kpc x 4 kpc (where -3 kpc$<x<$1 kpc and 2 kpc $< y < $6 kpc assuming a Cartesian grid) the same as in Fig.~2. All the panels correspond to a time of 200 Myr. In both simulations the molecular gas is situated in highly structured clumps. With a larger component of warm gas, there is more inter-arm molecular gas, which is sheared into filamentary clumps.}
\end{figure*}
   
\subsection{Molecular gas in the disk}
We show the distribution of molecular gas, overplotted on the total gas density in Fig.~6, for Runs A and B. Similar to previous results with a grand design potential \citep{Dobbs2007}, there is much more molecular gas when there are equal amounts of cold and warm gas, compared to when the gas is nearly all cold. This is because, as mentioned in the previous paper, more of the inter-arm gas is molecular, when it is confined to higher densities by the warm gas. There is also more molecular gas in these simulations compared to the grand design case. This is due to a combination of factors: a stronger potential, the fact that gas tends to remain in the spiral arms and collisions between spiral arms leading to more gas at higher densities. After 200 Myr (Table~1), 40\% and 23\% of the total disk mass is molecular for Runs A and B. However the system is not yet in equlibrium, and both fractions are still increasing.   

\subsection{Molecular clouds}
We also applied a clump-finding algorithm to the distribution of molecular gas in Runs A and B, predominantly to calculate the mass spectrum for comparison with previous results for grand design galaxies \citep{Dobbs2007}. For Run B, from Fig.~6, it is apparent that clumps in the outer part of the disk, where there is no spiral perturbation, have formed which are numerical in origin. At later times in the simulations, these clumps have retained high densities for a sufficient time that they contain molecular gas. We therefore only consider a region of 10 kpc by 10 kpc centred at the midpoint of the disk, which does not include these clumps. 

As for \citet{Dobbs2007}, we use a clump-finding algorithm to identify molecular clouds. We take the clump-finding method CF1 previously described in \citet{Dobbs2007}, which selects grid cells above a certain column density and classes groups of adjacent cells as a single clump.
We select cells which have a molecular gas surface density above 4 M$_{\odot}$ pc $^{-2}$, as used for our previous results \citep{Dobbs2007}, and only include clouds with $>$ 30 particles. The resolution of the simulations shown here is over twice that of previous analysis \citep{Dobbs2007} with the mass of each particle approximately 40 M$_{\odot}$.
The minimum total mass of a cloud is then $10^3$ M$_{\odot}$, although the mass of molecular gas may be less.

We find that the most massive molecular clouds ($> 10^5$ M$_{\odot}$)  are located where spiral arms are colliding or merging. This agrees with the previous predictions by \citet{Clarke2006} that enhanced star formation would occur here. We also find that there are very few inter-arm clouds between 2 and 4 kpc where most of the gas is in the spiral arms, whereas at lower radii there are more interarm clouds as a result of multiple collisions between spiral arms. 

Fig.~7 shows the mass spectra for Runs A and B. Assuming a mass spectrum of $dN/dM=M^{-\alpha}$, 
$\alpha=1.75\pm0.12$ for the simulation with predominantly cold gas, and $\alpha=2.05\pm0.1$ with a larger (50\%) component of hot gas. The mass distribution arises through cloud coagulation (e.g. \citealt{Field1965,Taff1972,Handbury1977,Scoville1979,Kwan1987,Das1996,Clark2006,Harfst2006}), a process enhanced by collisions between spiral arms. For previous simulations of grand design galaxies \citep{Dobbs2007}, we obtained similar values and also found that the mass spectrum for molecular clouds was steeper with a larger component of warm gas. This suggests that the nature of the ISM is more important in determining the mass spectrum than the nature of the potential, and how clump agglomeration occurs.
Previous observations of the Milky Way found $\alpha=1.5-1.8$ \citep{Solomon1987}, although some more recent surveys of external galaxies \citep{Blitz2007,Lada2007} indicate that $\alpha$ may be higher. 
\begin{figure}
\centerline{
\includegraphics[width=90mm]{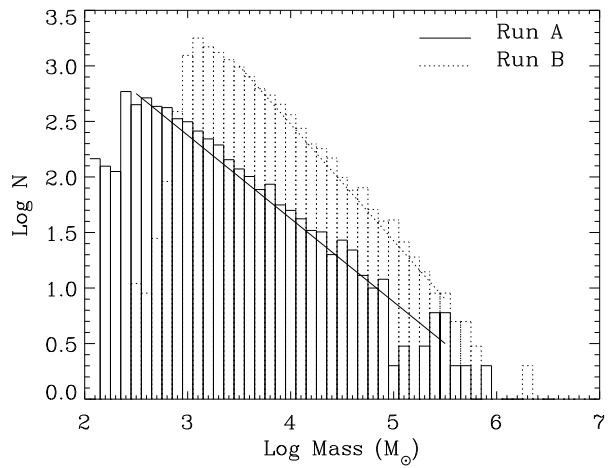}}
\caption{The clump mass spectra are shown for Runs A and B. Both are determined using the molecular mass of the clumps, after 200 Myr. The slopes indicated correspond to $\alpha=1.75$ (Run A) and $\alpha=2.05$ (Run B) where $\alpha$ is the index of the mass spectrum. The mass spectrum is steeper when more warm gas is present.}
\end{figure}

\section{Conclusion}
We have extended previous analysis of the structure and dynamics of the ISM in spiral galaxies by applying a time-dependent spiral potential derived from an N-body calculation. The gas forms long spiral arms, exhibiting a multi-armed structure where the underlying stellar component is determined by gravitational instabilities rather than a spiral density wave. The main difference compared to the grand design case is that the development, collision and dissipation of spiral arms in the underlying potential is  the main driver for generating high density structure in the ISM. The timescale for the evolution of structure is thus linked to the lifetime of the local minimum.

Gas accumulates in the potential minima as they develop, thus the densest gas is coincident with the stellar spiral arms. Consequently no offset would be expected between the dust lanes and star formation in such a galaxy.
Spurs associated with the shearing of GMCs as they leave the spiral arm, are largely absent with the active potential. This is because gas only leaves the spiral arms once the potential minimum dissipates. The gas then retains the large scale spiral arm structure, becoming slowly wound up until it fully dissipates or encounters another spiral arm. Instead of spurs, much larger bridges, which are remnants of former spiral arms, extend between the main spiral arms. 

We further identified molecular clouds from these simulations, finding that the largest GMCs occur where spiral arms are colliding. We found the surprising result that the mass spectra are similar to previous results where the potential corresponded to a density wave. This suggests that the external pressure applied to the cold gas determines the sizes and masses of the clouds, with a steeper index when the cold clumps are pressure confined.

A natural extension of this work would be to model both the stellar and gaseous components of the disc, to include the gravitational instabilities of the stars and the response of the gas simultaneously \citep{Li2005b}. Furthermore incorporating stellar feedback is required to treat the stellar and gaseous components self consistently. We have also neglected heating and cooling, magnetic fields and self gravity, which we are in the process of analysing with respect to grand design galaxies. We leave the consideration of these processes for more general stellar potentials to future work.  

\section*{Acknowledgments}
We are grateful to Jerry Sellwood and Dave Gittins for providing data from the N-body simulation described in Section~2.1.
We also thank the referee, Ant Whitworth, for helpful comments which significantly improved the clarity of the paper.
Computations included in this paper were performed using the UK Astrophysical
Fluids Facility (UKAFF). 

This work, conducted as part of the award `The formation of stars and planets: Radiation hydrodynamical and magnetohydrodynamical simulations' made under the European Heads of Research Councils and European Science Foundation EURYI (European Young Investigator) Awards scheme, was supported by funds from the Participating Organisations of EURYI and the EC Sixth Framework Programme. 

\bibliographystyle{mn2e}
\bibliography{Dobbs}

\bsp

\label{lastpage}

\end{document}